\documentclass{sig-alternate}
\usepackage{url}
\usepackage{amsmath}
\usepackage{amssymb}
\usepackage{bm}
\usepackage{stmaryrd}
\usepackage{color}
\usepackage{colortbl}
\usepackage{epsf}

\newcommand{\boxtheorem}{\textcolor{white}{abc} \hspace{-4mm} \hfill $\blacksquare$}
\newcommand{\ignore}[1]{}
\newcommand{\nit}[1]{{\it #1}}
\newcommand{\Q}{\mathcal{Q}}

\DeclareMathAlphabet{\mathpzc}{OT1}{pzc}{m}{it}
\newcommand{\match}{{\mathpzc m}}
\newcommand{\lub}{{\it lub}}
\newcommand{\glb}{{\it glb}}

\newcommand{\eat}[1]{}
\newcommand{\mc}[1]{\mathcal{ #1}}
\newcommand{\rlh}{\rightleftharpoons}

\newtheorem{proposition}{Proposition}
\newtheorem{lemma}{Lemma}

\newtheorem{theorem}{Theorem}
\newtheorem{definition}{Definition}

\newtheorem{example}{Example}

\newcommand{\red}[1]{\textcolor{red}{#1}}

\newcommand{\defproof}[2]{{\noindent\bf Proof of #1:\
}#2 \boxtheorem\\}

\newcommand{\comlb}[1]{{\vspace{2mm}\noindent \bf \red{COMM(LEO):}}~ #1 \hfill {\bf END.}\\}
\newcommand{\comsk}[1]{{\vspace{2mm}\noindent \bf \red{COMM(SOLMAZ):}}~ #1 \hfill {\bf END.}\\}
\newcommand{\comlaks}[1]{{\vspace{2mm}\noindent \bf \red{COMM(Laks):}}~ #1 \hfill {\bf END.}\\}

\title{Data Cleaning and Query Answering with Matching Dependencies and Matching Functions\thanks{Research funded by the BIN NSERC Strategic Network on BI (ADC01) and NSERC/IBM CRDPJ/371084-2008.}}

\numberofauthors{3}
\author{
\alignauthor
Leopoldo Bertossi\thanks{Faculty Fellow of the IBM CAS.  Also affiliated to University of Concepci\'{o}n (Chile).}\\
\affaddr{Carleton University} \\
\affaddr{Ottawa, Canada} \\
\email{bertossi@scs.carleton.ca}
\alignauthor
Solmaz Kolahi \\
\affaddr{University of British Columbia} \\
\affaddr{Vancouver, Canada} \\
\email{solmaz@cs.ubc.ca}
\alignauthor
Laks V. S. Lakshmanan \\
\affaddr{University of British Columbia} \\
\affaddr{Vancouver, Canada} \\
\email{laks@cs.ubc.ca}
}

\begin{document}
\maketitle

\ignore{
\begin{abstract}
Matching dependencies were recently introduced to specify record matching rules for
data cleaning and entity resolution purposes. Enforcing a matching dependency on a
database instance means that the value of some attributes for two tuples need to be
made identical provided that the value of some other attributes are sufficiently similar.
We formally introduce the process of cleaning an instance using matching dependencies
as a chase-like procedure, assuming that for every attribute a matching function is
available to decide which new value is to replace the two values that need to be made
equal. We show that matching functions naturally introduce a partial order (a lattice
structure to be more precise) on instances that shows semantic domination.
Using this partial order, we define the semantics of clean query answering by defining
certain/possible answers as the greatest lower bound/least upper bound of all possible
answers we can obtain after cleaning the instance. We show that clean query answering
is intractable in some cases. Then we study queries that behave monotonically wrt semantic
domination order, and show that we can provide an under/over approximation for clean
answers of monotone queries. More over, non-monotone positive queries can be relaxed into
a monotone, although less precise, query.
\end{abstract}}

\begin{abstract}
Matching dependencies were recently introduced as
declarative rules for
data cleaning and entity resolution. Enforcing a matching dependency on a
database instance identifies the values of some attributes for two tuples, provided that the values of
some other attributes are sufficiently similar. Assuming the existence of matching functions
for making two attributes values equal, we formally introduce the process of cleaning an instance using
matching dependencies,
as a chase-like procedure. We show that matching functions naturally introduce a lattice
structure on attribute domains, and a partial order of semantic domination between instances.
Using the latter, we define the semantics of clean query answering in terms of
certain/possible answers as the greatest lower bound/least upper bound of all possible
answers obtained from the clean instances. We show that clean query answering
is intractable in some cases. Then we study queries that behave monotonically wrt semantic
domination order, and show that we can provide an under/over approximation for clean
answers to monotone queries. Moreover, non-monotone positive queries can be relaxed into
monotone queries.
\end{abstract}

\section{Introduction}

Matching dependencies (MDs) in relational databases were recently 
introduced in \cite{Fan08} as a means of codifying a domain expert's
knowledge that is used in improving data quality. They specify that
a pair of attribute values in two database tuples are to be matched, i.e., made equal,
if similarities hold between other pairs of values in the same tuples. This is a generalization of
entity resolution \cite{elmargamid}, where basically  full tuples have to be merged or
identified since they seem to refer to the same entity of the outside reality.
This form of data fusion \cite{naumannACMCS} is important
in data quality assessment and in data cleaning.

Matching dependencies were formally studied
in \cite{FanJLM09}, as semantic constraints
for data cleaning and were given a model-theoretic semantics.
The main emphasis in that paper was on the
problem of entailment of MDs and on the existence of a formal axiom system for that task.

MDs as presented in \cite{FanJLM09} do not specify
\emph{how} the matching of attribute values is to be
done. In data cleaning, the user, on the basis of his or her experience and
knowledge of the application domain, may have a particular methodology or
heuristics for enforcing the identifications. \emph{In this paper we investigate
MDs in the context of matching functions}. These are functions that abstract the implementation
of value identification. Rather than investigate specific matching functions, we explore a
class of matching functions satisfying certain natural and intuitive axioms.
With these axioms, matching functions impose  a lattice-theoretic structure on
attribute domains. Intuitively, given two input attribute values that need
to be made equal, the matching function produces a value that {\em contains}
the information present in the two inputs and {\em semantically dominates} them.
We show this semantic domination partial order can be naturally lifted to
tuples of values as well as database instances as sets of tuples.

\begin{example} \em
Consider the following database instance $D_0$. Assume there is a matching dependency
stating that if for two tuples the values of name and phone are similar, then the value of address
should be made identical. Consider a similarity relation that indicates
the values of name and phone are similar for the two tuples in this instance.
To enforce the matching dependency, we create another instance $D_1$ in which the
value of address for two tuples is the result of applying a matching function
on the two previous addresses. This function combines the information in those
address values.
\begin{center}
\begin{tabular}{c|l|l|l}
$D_0$ & ${\it name}$ & ${\it phone}$ & ${\it address}$  \\ \hline
& John Doe & (613)123 4567 & Main St., Ottawa \\
& J. Doe & 123 4567 & 25 Main St. \\
\end{tabular} \ \\
{\large $\Downarrow$} \ \\
\begin{tabular}{c|l|l|l}
$D_1$ & ${\it name}$ & ${\it phone}$ & ${\it address}$  \\ \hline
& John Doe & (613)123 4567 & 25 Main St., Ottawa \\
& J. Doe & 123 4567 & 25 Main St., Ottawa \\
\end{tabular}
\end{center}
We can continue this process in a chase-like manner if there are still
other MD violations in $D_1$.
\boxtheorem
\end{example}
The framework of \cite{FanJLM09} leaves the implementation details of
data cleaning process with MDs completely unspecified and implicitly leaves it to the application on hand.
We point out some limitations of the proposal in \cite{FanJLM09} for purposes of cleaning dirty
instances in the presence of multiple MDs, and show that a formulation of the formal semantics of
the satisfaction and enforcement of MDs, incorporating matching functions, remedies this problem.
In giving such a formulation, we revisit the original semantics for MDs proposed in \cite{FanJLM09},
propose some changes and investigate their consequences. More precisely, we define \emph{intended
clean instances}, those that are obtained through the application of the MDs
in a chase-like procedure. We further investigate properties
of this procedure in relation to the properties of the matching functions, and show that,
in general, the chase procedure produces several different clean instances, each of which
semantically dominates the original dirty instance.

We then address the problem of query answering over a dirty instance,
where the MDs do \emph{not} hold. We take advantage of the semantic domination order
between instances, and define {\em clean answers} by specifying a tight lower bound
(corresponding to certain answers) and a tight upper bound (corresponding to possible answers) for
all answers that can be obtained from any of the possibly many clean instances.
We show that computing the exact bounds is intractable in general.
However, in polynomial time we can generate an under-approximation for certain answers as well
as an over-approximation for possible answers for queries that behave {\em monotonically}
w.r.t. the semantic domination order.

We argue that monotone queries provide more informative answers on instances that
have been cleaned with MDs and matching functions. We therefore introduce new
relational algebra operators that make use of the underlying lattice structure on
the domain of attribute values. These operators can be used to {\em relax} a regular
positive relational algebra query and make it monotone w.r.t. the semantic domination order.

Recently, Swoosh \cite{BenjellounGMSWW09} has been proposed as generic framework for
entity resolution. In entity resolution, whole
tuples are identified, or merged into a new tuple, whenever similarities hold between the tuples
on some attributes. Accordingly, the similarity and matching functions work
at the tuple level. Given their similarity of purpose, it is interesting to ask
what is the relationship between the frameworks of MDs and of Swoosh. We address this
question in this paper.

In summary, we make the following contributions:
\begin{itemize}
\item We identify the limitations of the original proposal of MDs \cite{FanJLM09}
wrt the application of data cleaning in the presence of multiple MDs and show
that the limitations can be overcome by considering MDs along with matching
functions.
\item We study matching functions in terms of their properties, which are
certain intuitive and natural axioms. Matching functions induce a lattice
framework on attribute domains which can be lifted to a partial order
over instances, that we call semantic domination.
\item We formally characterize answering a query given a dirty instance and a
set of MDs, and capture it using certain and possible answers. Computing these answers
is intractable in general. For queries that are monotone wrt the semantic
domination relation, we develop a polynomial time heuristic procedure for
obtaining under- and over-approximations of query answers.
\item We demonstrate the power of the framework of MDs and of our
chase procedure for MD application by reconstructing the
most common case for Swoosh, the so-called {\em union and merge} case,
in terms of matching dependencies with matching functions.
\end{itemize}
The paper is organized as follows. In Section~\ref{sec:back}, we provide necessary background on
matching dependencies as originally introduced. We introduce matching functions and the notion of
semantic domination in Section~\ref{sec:matFcs}. Then we define the data cleaning process with MDs in
Section~\ref{sec:enfor}. We explore the semantic of query answering in Section~\ref{sec:query}.
In Section~\ref{sec:monotone}, we study monotone queries and show how clean answers can be approximated.
We establish a connections to an important related work, Swoosh, in Section~\ref{sec:swoosh}, and present
concluding remarks in Section~\ref{sec:concl}.

\section{Background} \label{sec:back}
A database schema $\mathcal{R}$ is a set $\{R_1, \ldots, R_n\}$ of relation names.
Every relation $R_i$ is associated with a set of attributes, written as $R_i(A_1,\ldots,A_m)$, where
each attribute $A_j$ has a domain ${\it Dom}_{A_j}$. We assume that attribute names are
different across relations in the schema, but two attributes $A_j,A_k$ can be {\em comparable}, i.e.,
${\it Dom}_{A_j} = {\it Dom}_{A_k}$.
An instance $D$ of schema $\mathcal{R}$ assigns a finite set of tuples $t^D$ to every
relation $R_i$, where $t^D$ can be seen as a function that maps every attribute $A_j$ in $R_i$
to a value in ${\it Dom}_{A_j}$. We write $t^D[A_j]$ to refer to this value.
When $X$ is a list of attributes, we may write $t^D[X]$ to refer to the corresponding
list of attribute values. A tuple $t^D$ for a relation name $R \in \mc{R}$ is called an
$R$-tuple.
We deal with queries $\Q$ that are expressed in relational algebra, and treat them as operators
that map an instance $D$ to an instance $\Q(D)$.

For every attribute $A$ in the schema, we assume a binary similarity
relation $\approx_{\! A} \ \subseteq {\it Dom}_{\!A} \times {\it Dom}_{\!A}$.
Notice that whenever $A$ and $A'$ are comparable, the similarity relations
$\approx_A$ and $\approx_{A'}$ are identical.
We assume that each $\approx_A$ is symmetric and  subsumes
equality, i.e., $=_{\! \nit{Dom}_{\! A}} \ \subseteq \ \approx_{\! A}$.
When there is no confusion, we simply use $\approx$ for the similarity relation.
In particular, for lists of pairwise comparable attributes,  $X_i = A_1^i, \ldots, A_n^i$, $i=1,2$,
we write $X_1 \approx X_2$ to mean $A_1^1 \approx_1 A_1^2 \wedge  \cdots \wedge  A_n^1 \approx_n A_n^2$,
where $\approx_i$ is the similarity relation applicable to attributes $A_i^1, A_i^2$.

Given two pairs of pairwise comparable attribute lists $X_1, X_2$ and $Y_1, Y_2$ from
relations $R_1, R_2$, resp., a {\em matching dependency} (MD) \cite{FanJLM09} is a
sentence of the form
\begin{equation}
\varphi\!:  \ \ R_1[X_1] \approx R_2[X_2] \rightarrow R_1[Y_1] \rightleftharpoons R_2[Y_2].\footnote{All the variables
in $X_i, Y_j$ are implicitly universally quantified in front of the formula.}
\end{equation}
This dependency intuitively states that if for an $R_1$-tuple $t_1$ and an $R_2$-tuple $t_2$
in instance $D$, the attribute values in $t_1^D[X_1]$ are similar to attribute
values in $t_2^D[X_2]$, then we need to make the values $t_1^D[Y_1]$ and $t_2^D[Y_2]$
pairwise identical.

Enforcing MDs may cause a database instance $D$ to be changed to another instance $D'$.
To keep track of every single change, we assume that every tuple in an instance has a unique identifier $t$, which
could identify it in both instance $D$ and its changed version $D'$. We use $t^D$ and $t^{D'}$ to refer
to a tuple and its changed version in $D'$ that has resulted from enforcing an MD.
For convenience, we may use the terms tuple and tuple identifier interchangeably.

Fan et al.~\cite{FanJLM09} give a {\em dynamic semantics} for matching dependencies
in terms of a pair of instances: one where the similarities hold, and a second where
the specified identifications have been enforced:

\begin{definition}~ \em \cite{FanJLM09} \label{def:sat}
A pair of instances $(D,D')$ satisfies the MD $\varphi:
R_1[X_1] \approx R_2[X_2] \rightarrow R_1[Y_1] \rlh R_2[Y_2]$, denoted  $(D,D') \models \varphi$, if for
every $R_1$-tuple $t_1$ and $R_2$-tuple $t_2$ in $D$
that match the left-hand side of $\varphi$, i.e., $t_1^D[X_1] \approx t_2^D[X_2]$,
the following holds in the instance $D'$:
\begin{itemize}
\item [(a)] $t_1^{D'}[Y_1]=t_2^{D'}[Y_2]$, i.e., the values of the right-hand side attributes of $\varphi$ have been identified in $D'$; and
\item [(b)] $t_1,t_2$ in $D'$ match the left-hand side of $\varphi$, that is, $t_1^{D'}[X_1] \approx t_2^{D'}[X_2]$.
\end{itemize}
For a set $\Sigma$ of MDs, $(D,D') \models \Sigma$ iff $(D,D') \models \varphi$ for every $\varphi \in \Sigma$.
An instance $D'$ is called {\em stable} if $(D',D') \models \Sigma$. \boxtheorem
\end{definition}
Notice that a stable instance satisfies the MDs by itself, in the sense that all
the required identifications are already enforced in it. So, whenever we say that
an instance is {\em dirty}, we mean that it is not stable w.r.t. the given set of MDs.

While  this definition may be sufficient for the implication problem of MDs considered by
Fan et al.~\cite{FanJLM09}, it does not specify how a dirty database should be updated
to obtain a clean instance, especially when several {\em interacting updates} are required
in order to enforce all the MDs. Thus, it does not give a complete prescription for the purpose of
cleaning dirty instances. Moreover, from a different perspective,
the requirements in the definition may be too strong, as the following example shows.

\begin{example} \em   \label{ex:one} Consider the set of MDs $\Sigma$ consisting of \
$\varphi_1\!\!: R[A] \approx R[A] \rightarrow R[B] \rightleftharpoons R[B]$ and
$\varphi_2\!\!: R[B,C] \approx R[B,C] \rightarrow R[D] \rightleftharpoons R[D]$. The similarities are:
$a_1 \approx a_2$, $b_2 \approx b_3$, $c_2 \approx c_3$. Instance $D_0$ below is not a stable
instance, i.e., it does not satisfy $\varphi_1,\varphi_2$. We start by enforcing $\varphi_1$ on $D_0$.
\begin{center}
\begin{tabular}{c|cccc}
$D_0$ & $A$ & $B$ & $C$ & $D$ \\ \hline
& $a_1$ & $b_1$ & $c_1$ & $d_1$ \\
& $a_2$ & $b_2$ & $c_2$ & $d_2$ \\
& $a_3$ & $b_3$ & $c_3$ & $d_3$
\end{tabular} \hspace{2mm}
\begin{tabular}{c|cccc}
$D_1$ & $A$ & $B$ & $C$ & $D$ \\ \hline
& $a_1$ & $\langle b_1, b_2 \rangle $ & $c_1$ & $d_1$ \\
& $a_2$ & $\langle b_1, b_2 \rangle $ & $c_2$ & $d_2$ \\
& $a_3$ & $b_3$ & $c_3$ & $d_3$
\end{tabular}
\end{center}
Let $\langle b_1, b_2 \rangle$ in instance $D_1$ denote the value that replaces $b_1$ and $b_2$
to enforce $\varphi_1$ on instance $D_0$, and assume that $\langle b_1, b_2 \rangle \not \approx b_3$.
Now, $(D_0,D_1) \models \varphi_1$. However, $(D_0,D_1) \not \models \varphi_2$\ignore{ $(D_0,D_1) \not \models \Sigma$}.

If we identify $d_2, d_3$ via $\langle d_2,d_3\rangle$ producing instance $D_2$, the pair $(D_0,D_2)$
satisfies the condition (a) in Definition \ref{def:sat} for $\varphi_2$, but not condition (b). Notice that making more updates on $D_1$ (or $D_2$) to obtain
an instance $D'$, such that $(D_0,D') \models \Sigma$, seems hopeless as $\varphi_2$ will not
be satisfied because of the broken similarity that existed between $b_2$ and $b_3$. \boxtheorem
\end{example}
Definition \ref{def:sat} seems to capture well the one-step
enforcement of a single MD. However, as shown by the above example,
the definition has to be
refined in order to deal with sets of interacting MDs and to capture
an iterative process of MD enforcement. We address this problem in Section \ref{sec:enfor}.

Another issue worth mentioning is that  stable instances $D'$ for $D$ and $\Sigma$ are not subject to
any form of minimality criterion on  $D'$ in relation with $D$. We would expect such an instance
to be obtained via the enforcement of the MDs, without unnecessary changes. \ignore{The classical definition
of stable model semantics of logic programs does enjoy minimality and we would intuitively expect a
similar behavior of stability.} Unfortunately, this is not the case here: If in Example \ref{ex:one}
we keep only $\varphi_1$, and in instance $D_1$ we change $a_3$ by an arbitrary value $a_4$ that
is not similar to either $a_1$ or $a_2$, we obtain a stable instance with origin in $D_0$,
but the change of $a_3$ is unjustified and unnecessary. We will also address this issue.

Following~\cite{FanJLM09}, we assume in the rest of this paper that each MD is of the form
$R_1[X_1] \approx R_2[X_2] \rightarrow R_1[A_1] \rightleftharpoons R_2[A_2]$.
That is, the right-hand side of the MDs contains a pair of single attributes.
We also assume that the sets $\Sigma$ of MDs we consider are always finite.

\section{Matching Functions and Semantic Domination} \label{sec:matFcs}

In order to enforce a set of MDs (cf. Section \ref{sec:enfor}) we need an operation
that identifies two values whenever necessary. With  this purpose in mind, we will
assume that for each comparable pair of attributes $A_1,A_2$
with domain ${\it Dom}_{\!A}$, there is a binary {\em matching function}
$\match_A: {\it Dom}_{\!A}\times {\it Dom}_{\!A}\rightarrow {\it Dom}_{\!A}$, such that the value
$\match_A(a,a')$ is used to replace two values $a, a' \in {\it Dom}_{\!A}$ whenever the two values need to
be made equal. Here are a few natural properties to expect of the matching function
$\match_A$ (similar properties were considered in \cite{BenjellounGMSWW09},
cf. Section \ref{sec:swoosh}): For $a, a', a'' \in {\it Dom}_{\!A}$,
$$
\begin{array}{ll}
\text{{\bf \! I} (Idempotency):} & \! \! \! \match_A(a,a) = a,  \\
\text{{\bf C} (Commutativity):} & \! \! \! \match_A(a,a') = \match_A(a',a),  \\
\text{{\bf A} (Associativity):} & \! \! \! \match_A(a,\match_A(a',a'')) = \match_A(\match_A(a,a'), a''). \\
\end{array}
$$
It is reasonable to assume that any matching function satisfies at least these three axioms.
Under this assumption, the structure $({\it Dom}_{\!A}, \match_A)$  forms a {\em join semilattice}, $L_A$, that is,
a partial order with a  least upper bound ($\lub$) for every pair of elements. The induced
partial order $\preceq_A$ on the elements of ${\it Dom}_{\!A}$ is defined as follows: For every
$a, a' \in {\it Dom}_{\!A}$, $a \preceq_A a'$ whenever $\match_A(a,a') = a'$.
The {\em lub} operator with respect to this partial order
coincides with $\match_A$:  $\nit{lub}_{\preceq_A}\{a,a'\} = \match_A(a,a')$.

\ignore{
\comlb{What about a $\top$ element? It is mentioned at least in the following example.}
}

A natural interpretation for the partial order $\preceq_A$
in the context of data cleaning would be the notion of {\em semantic domination}. Basically,
for two elements $a,a' \in {\it Dom}_{\!A}$, we say that $a'$ {\em semantically dominates} \ $a$ if
we have $a \preceq_A a'$.
In the process of cleaning the data by enforcing matching dependencies,
we always replace two values $a,a'$, whenever certain similarities hold, by the value
$\match_A(a,a')$ that semantically dominates both $a$ and $a'$. This notion of domination is also related to
relative information contents \cite{BunemanJO91,scottTCS,imielinski84}.

To define the semantics of query answering on instances that have been cleaned with matching dependencies,
we might, in addition, need the existence of the greatest lower bound ($\glb$)
for any two elements in the domain
of an attribute. We therefore assume that $({\it Dom}_A, \match_A)$ is a lattice (i.e., both
$\lub$ and $\glb$ exist for every pair of elements in ${\it Dom}_A$ w.r.t. $\preceq_A$).
Moreover, there is a special element $\bot \in {\it Dom}_{\!A}$ such that $\match_A(a, \bot)= a$,
for every $a \in {\it Dom}_{\!A}$.
Notice that if we add an additional assumption to the semilattice, which requires that for every element
$a \in {\it Dom}_A$, the set $\{c \in {\it Dom}_A \mid c \preceq_A a\}$ (the set of elements $c$ with $\match_A(a,c)=a$),
is finite, then $\glb_{\!\preceq_A}\!\{a,a'\}$ does exist for every two elements $a,a' \in {\it Dom}_A$ and is equal to
$\lub_{\!\preceq_A}\!\{c \in {\it Dom}_A \mid c \preceq_A a \mbox{ and } c \preceq_A a'\}$.
We could also assume the existence of another special element
$\top \in {\it Dom}_{\!A}$
such that $\match_A(a, \top)= \top$, for every $a \in {\it Dom}_{\!A}$. This element could
represent the existence of inconsistency in data whenever matching dependencies force
to match two completely unrelated elements $a,a'$ from the domain, in which case
$\match_A(a,a') = \top$. However, \emph{the existence of $\top$ is not essential in our framework}.

\ignore{
As an alternative to assuming that matching functions with {\bf I,C,A} properties are given, we can start from
the assumption that the domain of each attribute $A$ forms a lattice w.r.t. some semantic domination partial
order $\preceq_A$, and then define matching functions as $\match_A(a,a') := \lub_{\preceq_A}\{a,a'\}$.
\comlaks{How is this different from what we do right now?}
}

\begin{example} \em
We give a few concrete examples of matching functions for different attribute domains. Our
example functions have all the properties {\bf I}, {\bf C}, and {\bf A}.
\begin{description}
\item[{\em Name, Address, Phone}]
Each atomic value $s$ of these string domains could be treated as a singleton set $\{s\}$.
Then a matching function $\match(S_1,S_2)$ for sets of strings $S_i$
could return $S_1 \cup S_2$. E.g., when matching addresses,
$\match(\mbox{\{`2366 Main Mall'\}}, \mbox{\{`Main Mall, Vancouver'\}})$ could return
$\{\mbox{`2366 Main Mall', `Main Mall, Vancouver'\}}$. \linebreak (This union matching function is further investigated
in Section \ref{sec:swoosh}.)
Alternatively, a more sophisticated matching function could merge two input strings into a third
string that contains both of the inputs. E.g., the match of the two input strings above could
instead be `2366 Main Mall, Vancouver'.
\item[{\em Age, Salary, Price}]
Each atomic value $a$ in these numerical domains could be treated as an interval $[a,a]$.
Then the matching function $\match([a_1,b_1], [a_2,b_2])$ would return the smallest
interval containing both $[a_1,b_1]$ and $[a_2,b_2]$, i.e., $\match([a_1,b_1], [a_2,b_2])=
[{\it min}\{a_1,a_2\}, {\it max}\{b_1,b_2\}]$.
\item[{\em Boolean Attributes}]
For attributes which take either a $0$ or $1$ value, the matching function would
return $\match(0,1) = \top$, where $\top$ shows inconsistency in the data,
and furthermore $\match(0,\top) = \top$ and $\match(1,\top)=\top$. In this case, the
purpose of applying the matching function is to \emph{record} the inconsistency in the
data and still conduct sound reasoning in presence of
inconsistency.\footnote{Matching of boolean attributes requires the existence of the
top element $\top$.} \hspace{8mm}$\blacksquare$
\end{description}

\end{example}
An additional property of matching functions worthy of consideration is
\emph{similarity preservation}, that is, the result of applying a matching function
preserves the similarity that existed between the old value being replaced and other
values in the domain. More formally:
$$
\begin{array}{ll}
\text{{\bf S} (Similarity Preservation):} & \! \! \! \text{ If } a \approx a' \text{, then } a \approx \match_A(a',a''), \\
\mbox{for every } a, a', a'' \in {\it Dom}_{\!A}.
\end{array}
$$
Unlike the previous properties ({\bf I}, {\bf C}, {\bf A}), property {\bf S} turns out  to be a
strong assumption, and we must consider both matching functions with {\bf S} and
without it.
Indeed, notice that since $\approx$ subsumes equality, and   $\match_A$ is commutative,
assumption \textbf{S} implies $a \approx \match_A(a,a')$ and $a' \approx \match_A(a,a')$, i.e.,
similarity preserving matching always results in a value similar to the value being replaced.
In the rest of the paper, we assume that for every comparable pair of attributes
$A_1,A_2$, there is an idempotent, commutative, and associative binary matching
function $\match_A$. Unless otherwise specified, we do not assume that these
functions are similarity preserving.

\begin{definition}  \em \label{def:domin}
Let $D_1,D_2$ be instances of schema ${\mathcal R}$, and  $t_1,t_2$ be two $R$-tuples
in $D_1,D_2$, respectively, with $R \in {\mathcal R}$. We write $t_1^{D_1} \preceq t_2^{D_2}$ if
$t_1^{D_1}[A] \preceq_A t_2^{D_2}[A]$ for every attribute $A$ in $R$.
We write $D_1 \sqsubseteq D_2$ if for every tuple $t_1$ in $D_1$, there is a tuple
$t_2$ in $D_2$, such that $t_1^{D_1} \preceq t_2^{D_2}$. \boxtheorem
\end{definition}
Clearly, the relation $\preceq$ on tuples can be applied to tuples in the same instance.
The ordering $\sqsubseteq$ on sets has been used in the context of complex objects~\cite{bancilhon89,KiferL89}
and also powerdomains, where it is called {\em Hoare ordering}~\cite{BunemanJO91}. It is also used
in \cite{BenjellounGMSWW09} for entity resolution (cf. Section \ref{sec:swoosh}).
It is known that for $\sqsubseteq$ to be a partial order, specifically to be antisymmetric,
we need to deal with {\em reduced} instances~\cite{bancilhon89}, i.e.,
instances $D$ in which there are no two different tuples $t_1,t_2$, such that $t_1^{D} \preceq t_2^{D}$.
We can obtain the {\em reduced version} of every instance $D$ by removing every tuple $t_1$, such
that $t_1^D \prec t_2^D$ for some tuple $t_2$ in $D$.

Next we will show that the set of reduced instances with the partial order $\sqsubseteq$ forms a
lattice: the least upper bound and the greatest lower bound for every finite set of reduced instances exist.
This result will be used later for query answering. We adapt some of the results from~\cite{bancilhon89},
where they prove a similar result for a lattice formed by the set of complex objects and the sub-object partial order.

\begin{definition} \em \label{def:theLat}
Let $D_1,D_2$ be instances of schema ${\mathcal R}$, and $t_1,t_2$
be two $R$-tuples in $D_1,D_2$, respectively, for $R \in {\mathcal R}$.
\begin{itemize}
\item [(a)]We define $D_1 \curlyvee D_2$ to be the reduced version of $D_1 \cup D_2$, where
$D_1 \cup D_2$ refers to the instance that takes the union of $R$-tuples from $D_1$ and $D_2$
for every  $R \in {\mathcal R}$.
\item [(b)]We define $t_1 \curlywedge t_2$ to be tuple $t$, such that $t[A] = \glb_{\preceq_A}\{t_1^{D_1}[A], t_2^{D_2}[A]\}$
for every attribute $A$ in $R$.
\item [(c)]We define $D_1 \curlywedge D_2$ to be the reduced version of the instance that assigns
the set of tuples $\{t_1 \curlywedge t_2 \mid t_1 \in  D_1, \ t_2 \in D_2, t_1, t_2 \ R\mbox{-tuples}\}$ to every
$R \in {\mathcal R}$. \boxtheorem
\end{itemize}
\end{definition}
Next we show that the operations defined in Definition~\ref{def:theLat} are equivalent to the
greatest lower bound and least upper bound of instances w.r.t. the partial order $\sqsubseteq$.

\begin{lemma} \label{lemma:lattice} \em
For every two instances $D_1,D_2$ and $R$-tuples $t_1,t_2$ in $D_1,D_2$, the
following holds:
\begin{enumerate}
\item $D_1 \curlyvee D_2$ is the least upper bound of $D_1,D_2$ w.r.t. $\sqsubseteq$.
\item $t_1 \curlywedge t_2$ is the greatest lower bound of $t_1,t_2$ w.r.t. $\preceq$.
\item $D_1 \curlywedge D_2$ is the greatest lower bound of $D_1,D_2$ w.r.t. $\sqsubseteq$. \boxtheorem
\end{enumerate}
\end{lemma}
In particular, we can see that $\preceq$ imposes a lattice structure on $R$-tuples.
Using Lemma~\ref{lemma:lattice}, we immediately obtain the following result.
\begin{theorem} \label{them:reduced} \em
The set of reduced instances for a given schema with the $\sqsubseteq$ ordering forms a lattice.
\boxtheorem
\end{theorem}

\section{Enforcement of MDs and Clean Instances }\label{sec:enfor}

In this section, we define {\em clean} instances that can be obtained from a dirty instance
by iteratively enforcing a set of MDs in a chase-like procedure.
Let $D,D'$ be two database instances with the same set of tuple identifiers, and
$t_1,t_2$ be an $R_1$-tuple and an $R_2$-tuple, respectively, in both $D$ and $D'$.
Let $\Sigma$ be a set of MDs, and
$\varphi: R_1[X_1] \approx R_2[X_2] \rightarrow R_1[A_1] \rightleftharpoons R_2[A_2]$ be
an MD in $\Sigma$.

\begin{definition}
\label{def:immediate} \em
Instance $D'$ is the immediate result of enforcing $\varphi$ on $t_1,t_2$ in
instance $D$, denoted by $(D,D')_{[t_1,t_2]} \models \varphi$, if
\begin{enumerate}
\item $t_1^D[X_1] \approx t_2^D[X_2]$, but $t_1^D[A_1] \neq t_2^D[A_2]$;
\item $t_1^{D'}[A_1] = t_2^{D'}[A_2] = \match_A(t_1^D[A_1], t_2^D[A_2])$; and
\item $D,D'$ agree on every other tuple and attribute value.
\end{enumerate} \boxtheorem
\end{definition}
Definition~\ref{def:immediate} captures a single step in a chase-like procedure that starts from
a dirty instance $D_0$ and enforces MDs step by step, by applying matching functions, until the
instance becomes stable. We propose that the output of this chase should be treated
as a clean version of the original instance, given a set of MDs, formally defined as follows.

\begin{definition}
\label{clean-def} \em
For an instance $D_0$ and a set of MDs $\Sigma$, an instance $D_k$ is
{\em $(D_0,\Sigma)$-clean} if $D_k$ is stable,
and there exists a finite sequence of
instances $D_1,\ldots, D_{k-1}$ such that, for every $i \in [1,k]$,
$(D_{i-1},D_i)_{[t_1^i,t_2^i]} \models \varphi^i$, for some $\varphi^i \in \Sigma$ and tuple
identifiers $t_1^i,t_2^i$. \boxtheorem
\end{definition}
Notice that if $(D_0,D_0) \models \Sigma$, i.e., it is already stable, then $D_0$ is its
only $(D_0,\Sigma)$-clean instance. Moreover, we have $D_{i-1} \sqsubseteq D_i$, for
every $i \in [1,k]$, since we are using matching functions to identify values. In particular,
we have $D_0 \sqsubseteq D_k$. In other words, clean instance $D_k$ semantically dominates
dirty instance $D_0$, and we might say $D_k$ it is {\em more informative} than $D_0$.

\begin{theorem} \em
\label{finite-prop}
Let $\Sigma$ be a set of matching dependencies and $D_0$ be an instance.
Then every sequence $D_1, D_2, \ldots $ such that $(D_{i-1},D_i)_{[t_1^i,t_2^i]} \models \varphi^i$,
for some $\varphi^i \in \Sigma$ and tuple identifiers $t_1^i,t_2^i$ in $D_{i-1}$, is
finite and computes a $(D_0,\Sigma)$-clean instance $D_k$ in polynomial number of steps in the size of $D_0$. \boxtheorem
\end{theorem}
In other words, the sequence of instances obtained by chasing MDs
reaches a fixpoint after polynomial number of steps.
That is, it is not possible to generate a new instance, because condition 1 in
Definition \ref{def:immediate}
is not satisfied by the last generated instance, i.e., the last instance is stable w.r.t. all MDs.
This is the consequence of assuming that matching functions are idempotent, commutative, and
associative.

Observe that, for a given instance $D_0$ and set of MDs $\Sigma$,
multiple clean instances may exist, each resulting from a different order of
application of  MDs on $D_0$ and from different selections of violating tuples.
It is easy to show that the number of clean instances is finite.

Notice also that  for a $(D_0,\Sigma)$-clean instance $D_k$, we may have
$(D_0,D_k) \not \models \Sigma$ (cf. Definition \ref{def:sat}).
Intuitively, the reason is that some
of the similarities that existed in $D_0$ could have been broken by iteratively enforcing the MDs to
produce $D_k$.  We argue that this is a price we may have to pay if we want to enforce a set of
interacting MDs. However, each $(D_0,\Sigma)$-clean instance is stable and captures the persistence of attribute values
that are not affected by MDs.
The following example illustrates these points. For simplicity, we write
$\langle a_1,\ldots,a_l \rangle$ to represent $\match_A(a_1,\match_A(a_2,\match_A(\ldots,a_l)))$,
which is allowed by the associativity assumption.

\begin{example} \em
\label{nocycle-example} Consider the set of MDs $\Sigma$ consisting of \
$\varphi_1\!\!: R[A] \approx R[A] \rightarrow R[B] \rightleftharpoons R[B]$ and
$\varphi_2\!\!: R[B] \approx R[B] \rightarrow R[C] \rightleftharpoons R[C]$.
We have the similarities: $a_1 \approx a_2$, $b_2 \approx b_3$. The following
sequence of instances leads to a $(D_0,\Sigma)$-clean instance $D_2$. \ \\ \ \\
\begin{tabular}{c|ccc}
$D_0$ & $A$ & $B$ & $C$ \\ \hline
& $a_1$ & $b_1$ & $c_1$ \\
& $a_2$ & $b_2$ & $c_2$ \\
& $a_3$ & $b_3$ & $c_3$
\end{tabular} \hspace{0.1cm}
\begin{tabular}{c|ccc}
$D_1$ & $A$ & $B$ & $C$ \\ \hline
& $a_1$ & $\langle b_1,b_2 \rangle$ & $c_1$ \\
& $a_2$ & $\langle b_1,b_2 \rangle$ & $c_2$ \\
& $a_3$ & $b_3$ & $c_3$
\end{tabular} \\ \ \\
\begin{tabular}{c|ccc}
$D_2$ & $A$ & $B$ & $C$ \\ \hline
& $a_1$ & $\langle b_1,b_2 \rangle$ & $\langle c_1,c_2 \rangle$ \\
& $a_2$ & $\langle b_1,b_2 \rangle$ & $\langle c_1,c_2 \rangle$ \\
& $a_3$ & $b_3$ & $c_3$
\end{tabular}
\ \\ \ \\
However, $(D_0,D_2) \not \models \Sigma$, and
the reason is that $\langle b_1,b_2 \rangle \approx b_3$ does not necessarily hold.
We can enforce the MDs in another order and obtain a different $(D_0,\Sigma)$-clean instance: \ \\ \ \\
\begin{tabular}{c|ccc}
$D_0$ & $A$ & $B$ & $C$ \\ \hline
& $a_1$ & $b_1$ & $c_1$ \\
& $a_2$ & $b_2$ & $c_2$ \\
& $a_3$ & $b_3$ & $c_3$
\end{tabular} \hspace{0.1cm}
\begin{tabular}{c|ccc}
$D'_1$ & $A$ & $B$ & $C$ \\ \hline
& $a_1$ & $b_1$ & $c_1$ \\
& $a_2$ & $b_2$ & $\langle c_2, c_3 \rangle$ \\
& $a_3$ & $b_3$ & $\langle c_2, c_3 \rangle$
\end{tabular} \\ 
\begin{tabular}{c|ccc}
$D'_2$ & $A$ & $B$ & $C$ \\ \hline
& $a_1$ & $\langle b_1, b_2 \rangle$ & $c_1$ \\
& $a_2$ & $\langle b_1, b_2 \rangle$ & $\langle c_2, c_3 \rangle$ \\
& $a_3$ & $b_3$ & $\langle c_2, c_3 \rangle$
\end{tabular} \hspace{0.1cm}
\begin{tabular}{c|ccc}
$D'_3$&$A$&$B$&$C$\\ \hline
&$a_1$&$\langle b_1, b_2 \rangle$ & $\langle c_1, c_2, c_3 \rangle$\\
& $a_2$ & $\langle b_1, b_2 \rangle$ & $\langle c_1, c_2, c_3 \rangle$\\
& $a_3$ & $b_3$ & $\langle c_2, c_3 \rangle$
\end{tabular} \hspace{0.1cm}
\ \\
Again, $D'_3$ is a $(D_0,\Sigma)$-clean instance, but $(D_0, D'_3) \not \models \Sigma$.
\boxtheorem
\end{example}
It would be interesting to know when there is only one
$(D_0,\Sigma)$-clean instance $D_k$, and also when, for a clean instance $D_k$,
$(D_0, D_k) \models \Sigma$ holds. The following two propositions establish
natural sufficient conditions for these properties to hold.
\ignore{
result shows that
these properties hold if the matching functions for all attributes are similarity preserving.
}

\begin{proposition} \em \label{prop:unique}
Suppose that for every pair of comparable attributes $A_1,A_2$,
the matching function $\match_A$ is similarity preserving. Then, for every
set of MDs $\Sigma$ and every instance $D_0$, there is a unique
$(D_0,\Sigma)$-clean instance $D_k$. Furthermore,  $(D_0, D_k) \models \Sigma$. \boxtheorem
\end{proposition}
We say that a set of matching dependencies $\Sigma$ is {\em interaction-free} if
for every two MDs $\varphi_1, \varphi_2 \in \Sigma$, the two sets of attributes on the
right-hand side of $\varphi_1$ and left-hand side of $\varphi_2$ are disjoint. The two sets
of MDs in Examples \ref{ex:one} and \ref{nocycle-example} are not interaction-free.

\begin{proposition} \em \label{prop:nointeraction}
Let $\Sigma$ be an interaction-free set of MDs. Then for every instance
$D_0$, there is a unique $(D_0,\Sigma)$-clean instance $D_k$.
Furthermore,  $(D_0, D_k) \models \Sigma$. \boxtheorem
\end{proposition}
The chase-like procedure that produces a $(D_0,\Sigma)$-clean instance
makes only those changes to instance $D_0$ that are necessary, and are
imposed by the dynamic semantics of MDs.
In this sense, we can say that the chase implements minimal changes necessary to
obtain a clean instance.

Another interesting question is whether $(D_0,\Sigma)$-clean instances
are at a minimal distance to $D_0$ w.r.t. the partial order $\sqsubseteq$.
This is not true in general. For instance in Example~\ref{nocycle-example},
observe that for the two $(D_0,\Sigma)$-clean instances $D_2$ and $D'_3$,
$D_2 \sqsubseteq D'_3$, but $D'_3 \not \sqsubseteq D_2$,
which means $D'_3$ is not at a minimal distance to $D_0$ w.r.t. $\sqsubseteq$.
However, both of these clean instances may be useful in query answering, because,
informally speaking, they can provide a lower bound and an upper bound for the
possible clean interpretations of the original dirty instance w.r.t. the semantic domination.
This issue is discussed in the next section.

\ignore{
\comlb{Is it possible to characterize the $(D_0,\Sigma)$-clean instances as the stable instances
that ``minimally" differ from $D_0$? Possibly related to $\sqsubseteq$? I remember we discussed this.
Did we?}
\comsk{Yes, we explored this, but it's not true. There are $(D_0,\Sigma)$-clean instances that are
not $\sqsubseteq$-minimally different from $D_0$. Also, there are stable instances that
are $\sqsubseteq$-minimally different from $D_0$, but they are not $(D_0,\Sigma)$-clean (can't
be obtained by chase). I have counter examples.}

\comlb{NEW: This could be mentioned in the Discussion/Conclusions section, with the counterexample.
It is a natural question.}
}

\section{Clean Query Answering} \label{sec:query}

Most of the literature on data cleaning has concentrated on producing a clean instance starting from
a dirty one. However, the problem of characterizing and retrieving the data in the original instance
that can be considered to be clean has been neglected. In this section we study  this problem, focusing on
query answering. More precisely,
given an instance $D$, a set $\Sigma$ of MDs, and a query 
$\Q$ posed to $D$, we want to characterize the answers that are consistent with $\Sigma$, i.e.,
that would be returned by an instance where all the MDs have been enforced. Of course, we have
to take into account that there may be several such instances.

This situation is similar to the one
encountered in {\em consistent query answering} (CQA) \cite{ABC99,Ber06,chom07}, where query answering
is characterized and performed on database instances that may fail to satisfy certain classic
integrity constraints (ICs). For such a database instance, a {\em repair} is an instance that satisfies
the integrity constraints and minimally differs from the original instance. For a given query,
a {\em consistent answer} (a form of certain answer) is defined as the set of tuples that are present in
the intersection of answers to the query when posed to every repair.
A less popular alternative is the notion of {\em possible answer}, that is defined as the union of
all tuples that are present in the answer to the query when posed to every repair.

A similar semantics for clean query answering under matching
dependencies can be defined. However,  the partial order
relationship $\sqsubseteq$ between a dirty instance and its clean instances establishes an
important difference between clean instances w.r.t. matching
dependencies and repairs w.r.t. traditional ICs.

Intuitively, a clean instance has improved the information that already existed in the dirty
instance and made it more informative and consistent. We would like to carefully take
advantage of this partial order relationship and use it in the definition of certain and
possible answers. We do this by taking the greatest lower bound (glb) and least upper bound (lub) of answers
of the query over multiple clean instances, instead of taking the set-theoretic intersection
and union.

Let $\Sigma$ be a set of MDs, $D_0$ be a database instance, and $\Q$ be a query posed to
instance $D_0$. We define {\em certain} and {\em possible answers} as follows.
\begin{eqnarray}
{\it Cert}_{\Q}(D_0) = \glb_\sqsubseteq \{ \Q(D) \mid D \text{ is a } (D_0,\Sigma)\text{-clean instance}\}.
\label{eq:cert}\\
{\it Poss}_{\Q}(D_0)= \lub_\sqsubseteq \{ \Q(D) \mid D \text{ is a } (D_0,\Sigma)\text{-clean instance}\}.
\label{eq:poss}
\end{eqnarray}
The glb and lub above are defined on the basis of the partial order $\sqsubseteq$ on sets of tuples. Since
there is a finite number of clean instances for $D_0$, these glb and  lub exist (cf. Theorem \ref{them:reduced}). In Eq.
 (\ref{eq:cert}) and (\ref{eq:poss}) we are assuming that
each of the $\Q(D)$ is reduced (cf. Section~\ref{sec:matFcs}). By Definition \ref{def:theLat}, ${\it Cert}_{\Q}(D_0)$
and ${\it Poss}_{\Q}(D_0)$ are also reduced. Moreover, we clearly have
${\it Cert}_{\Q}(D_0) \sqsubseteq {\it Poss}_{\Q}(D_0)$.

The following example motivates these choices. It also shows that, unlike some cases of inconsistent
databases and consistent query answering, certain answers could be quite informative
and meaningful for databases with matching dependencies.

\begin{example} \em
\label{address-ex}
Consider relation $R({\it name}, {\it phone}, {\it address})$, and
set $\Sigma$ consisting of the following MDs: \ignore{$\varphi_1, \varphi_2$, where}
$$
\begin{array}{ll}
\varphi_1: & R[{\it name}, {\it phone}, {\it address}] \approx R[{\it name}, {\it phone}, {\it address}] \rightarrow \\
& R[{\it address}] \rightleftharpoons R[{\it address}], \\
\varphi_2: & R[{\it phone}, {\it address}] \approx R[{\it phone}, {\it address}] \rightarrow \\
& R[{\it phone}] \rightleftharpoons R[{\it phone}].
\end{array}
$$
Suppose that in the dirty instance $D_0$, shown below, the following similarities hold:
$$
\begin{array}{l}
\text{``John Doe''} \approx \text{``J. Doe''}, \\
\text{``Jane Doe''} \approx \text{``J. Doe''}, \\
\text{``(613)123 4567''} \approx \text{``123 4567''}, \\
\text{``(604)123 4567''} \approx \text{``123 4567''}, \\
\text{``25 Main St.''} \approx \text{``Main St., Ottawa''}, \\
\text{``25 Main St.''} \approx \text{``25 Main St., Vancouver''}.
\end{array}
$$
Other non-trivial similarities that are not mentioned do not hold.
Moreover, the matching functions act as follows:
$$
\begin{array}{l}
\match_{phone}(\text{``(613)123 4567'', ``123 4567''}) = \text{``(613)123 4567''}, \\
\match_{phone}(\text{``123 4567'', ``(604)123 4567''}) = \text{``(604)123 4567''}, \\
\match_{address}(\text{``Main St., Ottawa'', ``25 Main St.''}) = \\ \hspace*{4.6cm}\text{``25 Main St., Ottawa''}, \\
\match_{address}(\text{``25 Main St.'', ``25 Main St., Vancouver''}) =\\ \hspace*{4.6cm} \text{``25 Main St., Vancouver''}.
\end{array}
$$

\begin{center}
\begin{tabular}{c|l|l|l}
$D_0$ & ${\it name}$ & ${\it phone}$ & ${\it address}$  \\ \hline
& John Doe & (613)123 4567 & Main St., Ottawa \\
& J. Doe & 123 4567 & 25 Main St. \\
& Jane Doe & (604)123 4567 & 25 Main St., Vancouver \\
\end{tabular} \ \\ \ \\ \ \\
\end{center}
Observe that from $D_0$ we can obtain
two different $(D_0,\Sigma)$-clean instances $D,D'$, depending on the order of
enforcing MDs.
\begin{center}
\begin{tabular}{c|l|l|l}
$D$ & ${\it name}$ & ${\it phone}$ & ${\it address}$  \\ \hline
& John Doe & (613)123 4567 & 25 Main St., Ottawa \\
& J. Doe & (613)123 4567 & 25 Main St., Ottawa \\
& Jane Doe & (604)123 4567 & 25 Main St., Vancouver \\
\end{tabular} \ \\ \ \\ \ \\

\begin{tabular}{c|l|l|l}
$D'$ & ${\it name}$ & ${\it phone}$ & ${\it address}$  \\ \hline
& John Doe & (613)123 4567 & Main St., Ottawa \\
& J. Doe & (604)123 4567 & 25 Main St., Vancouver \\
& Jane Doe & (604)123 4567 & 25 Main St., Vancouver \\
\end{tabular}
\end{center}
Now consider the query $\Q: \pi_{\text{address}} (\sigma_{\text{name=``J. Doe''}} R)$,
asking for the residential address of J. Doe.
We are interested in a certain answer.
It can be obtained by taking the greatest lower
bound of the two answer sets:
$$
\begin{array}{l}
\Q(D)  = \{(\mbox{``25 Main St., Ottawa"})\}, \\
\Q(D')  = \{(\mbox{``25 Main St., Vancouver"})\}.
\end{array}
$$
In this case, and according to \cite{bancilhon89}, and using Lemma \ref{lemma:lattice},
$$
\begin{array}{ll}
\nit{glb}_\sqsubseteq\{\Q(D),\Q(D')\} = \{a \curlywedge a'~|~ a \in \Q(D), \ a' \in \Q(D')\}  \\
= \{(\mbox{``25 Main St., Ottawa"}) \curlywedge (\mbox{``25 Main St., Vancouver"})\} \\
= \{(\glb_{\preceq_\nit{Address}}\{\mbox{``25 Main St., Ottawa"},\mbox{``25 Main St.,}\\ \hspace*{6.3cm}\mbox{Vancouver"}\})\} \\
= \{(\mbox{``25 Main St."})\}.
\end{array}
$$
We can see that, no matter how we clean $D_0$, we can say for sure that J. Doe is
at 25 Main St.
Notice that the set-theoretic intersection
of the two answer sets is empty.
If we were interested in all possible answers, we could take
the least upper bound of two answer sets, which would be the union of
the two in this case.
\boxtheorem
\end{example}
We define {\em clean answer} to be an upper and lower bound of query answers
over all possible clean interpretations of a dirty database instance. This definition is inspired by
the same kind of approximations used in the contexts of partial and incomplete
information~\cite{Lipski79,AbiteboulKG91},
inconsistent databases~\cite{ABC99,Ber06,chom07}, and data exchange~\cite{Libkin06}.
These upper and lower bounds could provide useful information about the value of
aggregate functions, such as sum and count \cite{ArenasBC03b,fux07,AK08}.

\ignore{
\comlb{ Do we have ${\it Cert}_{\Q}(D_0) \sqsubseteq {\it Poss}_{\Q}(D_0)$?}
\comsk{It should be easy to show that ${\it Cert}_{\Q}(D_0) \sqsubseteq {\it Poss}_{\Q}(D_0)$
for $\sqsubseteq$-monotone queries.
 We need to say something about how lub and glb are computed algorithmically,
which ends with eliminating redundant tuples.}
\comlb{At this stage we haven't talked about monotonicity, so we can postpone this for later in the paper.}
}

\begin{definition} \em
For a query $\Q$ posed to a database instance $D_0$ and a set of MDs $\Sigma$,
a {\em clean answer} is specified by two bounds as

\vspace{2mm}\hspace{8mm}
$
{\it Clean}_{\Q}(D_0) = ({\it Cert}_{\Q}(D_0), {\it Poss}_{\Q}(D_0)).
$ \boxtheorem
\end{definition}
Notice, from the results in Section \ref{sec:enfor}, that in the case of having similarity-preserving
matching functions or non-interacting matching dependencies, these bounds would collapse into a
single set, which is obtained by running the query on the unique clean instance.

\subsubsection*{Complexity of Computing Clean Answers}
Here we study the complexity of computing clean answers over database
instances in presence of MDs. As with incomplete and
inconsistent databases, this problem easily becomes intractable for simple MDs and queries,
which motivates the need for developing approximate solutions to these problems. We explore
approximate solutions for queries that behave monotonically w.r.t. the partial
order $\sqsubseteq$ in Section~\ref{approx-sec}.

\begin{theorem} \em \label{theo:coNP}
There are a schema with two interacting MDs and a relational algebra query, for which
deciding \linebreak whether a tuple belongs to the certain answer set for an instance
$D_0$ is coNP-complete (in the size of $D_0$).
\boxtheorem
\end{theorem}

\section{Monotone Queries} \label{sec:monotone}

So far we have seen that clean instances are a more informative view of a dirty instance
obtained by enforcing matching dependencies. That is, $D_0 \sqsubseteq D$, for every
$(D_0,\Sigma)$-clean instance $D$. From this perspective, it would be natural
to expect that for a positive query, we would obtain a more informative answer if we pose
it to a clean instance instead of to the dirty one. We can translate this requirement into
a monotonicity property for queries w.r.t. the partial order $\sqsubseteq$.

\begin{definition} \em
A query $\Q$ is $\sqsubseteq${\em -monotone} if, for every pair of instances $D,D'$,
such that $D \sqsubseteq D'$, we have $\Q(D) \sqsubseteq \Q(D')$. \boxtheorem
\end{definition}
Monotone queries have an
interesting behavior when computing clean answers. For these queries,
we can under-approximate (over-approximate) certain answers (possible answers)
by taking the greatest lower bound (least upper bound)
of all clean instances and then running the query on the result.
Notice that we are not claiming that these are polynomial-time approximations.

\begin{proposition} \em
\label{prop:monotone}
If $\mathcal{D}$ is a finite set of database instances and $\Q$ is a $\sqsubseteq$-monotone query,
the following holds:
\begin{equation} \label{glb-equation}
\Q(\glb_\sqsubseteq \{ D \mid D \in \mathcal{D} \}) \sqsubseteq \glb_\sqsubseteq \{ \Q(D) \mid D \in \mathcal{D} \},
\end{equation}
\begin{equation} \label{lub-equation}
\lub_\sqsubseteq \{ \Q(D) \mid D \in \mathcal{D} \} \sqsubseteq \Q(\lub_\sqsubseteq \{ D \mid D \in \mathcal{D} \}). 
\end{equation} \boxtheorem
\end{proposition}
As is well known, positive relational algebra queries composed of selection, projection,
Cartesian product, and union,
are monotone. However, the following example shows that monotonicity does not hold for
very simple positive queries involving selections.

\begin{example} \em
\label{not-monotone-ex}
Consider instance $D_0$ in Example~\ref{address-ex} and two $(D_0,\Sigma)$-clean instances
$D$ and $D'$. Let $\Q$ be a query asking for names of people residing at ``25 Main St.'',
expressed as relational algebra expression
$\pi_{{\it name}} (\sigma_{{\it address}=\text{``25 Main St.''}} (R))$.
Observe that $\Q(D_0) = \{(\text{``J. Doe''})\}$, and $\Q(D) = \Q(D')$ $= \emptyset$. Query $\Q$
is therefore not monotone, because we have $D_0 \sqsubseteq D$, $D_0 \sqsubseteq D'$,
but $\Q(D_0) \not \sqsubseteq \Q(D)$, $\Q(D_0) \not \sqsubseteq \Q(D')$.
\boxtheorem
\end{example}
It is not surprising that $\sqsubseteq$-monotonicity is not captured by usual relational queries,
in particular, by queries that \emph{are} monotone w.r.t. set inclusion.
After all, the queries we have considered
so far do not even mention the $\preceq$-lattice that is at the basis of the $\sqsubseteq$ order.
Next we will consider queries expressing conditions in term of the semantic domination lattice.

\subsection{Query relaxation}
As shown in Example~\ref{not-monotone-ex}, we may not get the answer we expect by running
a usual relational algebra query on an instance that has been cleaned using matching
dependencies. We therefore propose to {\em relax} the queries, by taking advantage of
the underlying $\preceq$-lattice structure obtained from matching functions, to make them
$\sqsubseteq$-monotone.
In this way, we achieve two goals: First, the resulting queries provide more informative answers;
and second, we can take advantage of Proposition~\ref{prop:monotone}
to approximate clean answers from below and from above.

We introduce the (negation free) language {\em relaxed relational algebra}, $\mc{RA}_{\preceq}$,
by providing two selection operators $\sigma_{a \preceq A}$
and $\sigma_{A_1 \Join_{\!\!\preceq} A_2}$ (for comparable attributes $A_1,A_2$), defined as
follows.

\begin{definition} \em The language $\mc{RA}_{\preceq}$  is composed of relational operators
$\pi, \times, \cup$ (with usual definitions), plus $\sigma_{a \preceq A}$, and
$\sigma_{A_1 \Join_{\preceq} A_2}$, defined by:
$$\begin{array}{l}
\sigma_{a \preceq A}(D) = \{ t^D \mid a \preceq_A t^D[A] \} \ \ \text{ (here } a \in \nit{Dom}_A \text{)}, \\
\sigma_{A_1 \Join_{\!\!\preceq} A_2}(D) = \{ t^D \mid \exists a \in {\it Dom}_A \text{ s.t. }
a \preceq_A t^D[A_1], \\ \hspace{2.9cm} \; a \preceq_A t^D[A_2], \; a \neq \bot \}. \hspace{1.9cm}\blacksquare
\end{array}$$
\end{definition}
For string attributes, for instance, the selection operator $\sigma_{a \preceq A}$ checks whether
the value of attribute $A$ dominates the substring $a$, and the join selection operator
$\sigma_{A_1 \Join_{\!\!\preceq} A_2}$ checks whether the values of attributes $A_1,A_2$
dominate a common substring.
Notice that queries in the language $\mc{RA}_{\preceq}$ are not domain independent:
The result of posing a query to an instance depends not only on the values in the active domain
of the instance but also on the domain lattices. In other words, query answering depends
on how data cleaning is being implemented.

It can be easily observed that all operators in the language $\mc{RA}_{\preceq}$
are $\sqsubseteq$-monotone, and therefore every query expression in $\mc{RA}_{\preceq}$
that is obtained by composing these operators is also $\sqsubseteq$-monotone.

\begin{proposition}
\label{prop:relaxedra} \em
Let $\Q$ be a query in $\mc{RA}_{\preceq}$. For every two instances $D,D'$
such that $D \sqsubseteq D'$, we have $\Q(D) \sqsubseteq \Q(D')$. \boxtheorem
\end{proposition}
\ignore{
\comlb{I have the feeling that the language just introduced may be more interesting than just
for query relaxation. It could be sold as a ``semantic-domination sensitive query language",
which may be interesting and natural on its own. If you agree with me, it could be moved to the
main body of this section (pushed up) or even have a subsection name of its own.}
\comsk{I agree, the language is interesting on its own. However, query relaxation is nothing
but introducing this language and we won't have enough results to make it two sections.}
}
Now suppose that we have a query $\Q$, written in positive relational algebra, i.e.,
composed of $\pi, \times, \cup, \sigma_{A=a}, \sigma_{A_1=A_2}$, the last two being {\em hard}
selection conditions, which is to be posed to an instance $D_0$. After cleaning
$D_0$ by enforcing a set of MDs $\Sigma$ to obtain a $(D_0,\Sigma)$-clean instance $D$,
running query $\Q$ on $D$ may no longer provide the expected answer, because
some of the values have changed in $D$, i.e., they have semantically grown w.r.t. $\preceq$.
In order to capture this semantic growth, our query relaxation framework proposes
the following {\em query rewriting} methodology: Given a query $\Q$ in positive RA, transform it into a
query $\Q_{\preceq}$ in $\mc{RA}_{\preceq}$ by simply replacing the selection operators
$\sigma_{A=a}$ and $\sigma_{A_1=A_2}$ by $\sigma_{a \preceq A}$
and $\sigma_{A_1 \Join_{\preceq} A_2}$, respectively.

\begin{example} \em
Consider again instance $D_0$ in Example~\ref{address-ex} and
$(D_0,\Sigma)$-clean instances $D$ and $D'$, and query $\Q$
asking for names of people residing at ``25 Main St.'', expressed as
$\pi_{{\it name}} (\sigma_{{\it address}=\text{``25 Main St.''}} (R))$.
We obtain the empty answer from each of $D, D'$. So, in this case the certain and the
possible answers are empty, a not very informative outcome.

However, after the relaxation rewriting of $\Q$, we obtain the query
$\Q_{\preceq}\!\!: \pi_{{\it name}}($$\sigma_{\text{``25 Main St.''} \preceq {\it address}} (R))$.
If we pose $\Q_{\preceq}$ to the clean instances, we obtain
$$
\begin{array}{l}
\Q_{\preceq} (D) = \{(\text{``John Doe''}), (\text{``J. Doe''}), (\text{``Jane Doe''})\}, \\
\Q_{\preceq} (D') = \{(\text{``J. Doe'')}, (\text{``Jane Doe''})\},
\end{array}
$$
and thus ${\it Cert}_{\Q_{\preceq}}(D_0)$ $= \{(\text{``J. Doe''}), (\text{``Jane Doe''})\}$.
This outcome is much more informative; and, above all, is sensitive to the underlying information
lattice.
\boxtheorem
\end{example}

\begin{proposition}
\label{prop:rewrite} \em
For every positive relational algebra query $\Q$ and every instance $D$,
we have $\Q(D) \sqsubseteq \Q_{\preceq}(D)$, where $\Q_{\preceq}$
is the relaxed rewriting of  $\Q$. \boxtheorem
\end{proposition}


\subsection{Approximating Clean Answers}
\label{approx-sec}
Given the high computational cost of clean query answering when there are
multiple clean instances, it would be desirable to provide
an approximation to clean answers that is computable in polynomial time.
In this section, we are interested in approximating clean answers by producing
an under-approximation of certain answers and an over-approximation of possible
answers for a given monotone query $\Q$. That is, we would like to obtain
$(\Q_{\downarrow}(D_0), \Q_{\uparrow}(D_0))$, such that
$\Q_{\downarrow}(D_0) \sqsubseteq {\it Cert}_{\Q}(D_0)$ and
${\it Poss}_{\Q}(D_0) \sqsubseteq \Q_{\uparrow}(D_0)$.

Since $\Q$ is a monotone query,  by Proposition~\ref{prop:monotone}, we have
$\Q(\glb_\sqsubseteq \{ D \mid D \mbox{ is } (D_0,\Sigma)\mbox{-clean}\})
\sqsubseteq {\it Cert}_{\Q}(D_0)$, and moreover,
${\it Poss}_{\Q}(D_0) \sqsubseteq \Q(\lub_\sqsubseteq \{ D \mid D \mbox{ is}$
$(D_0,\Sigma)\mbox{-clean}\})$.
In consequence, it is good enough to find under- and over-approximations for the greatest lower
bound and the least upper bound, resp., of the set of all $(D_0,\Sigma)$-clean instances,
and then pose $\Q$ to these approximations to obtain $\Q_{\downarrow}(D_0)$ and
$\Q_{\uparrow}(D_0)$.

The reason for having multiple clean instances is that matching dependencies are
not necessarily interaction-free and the matching functions are not necessarily
similarity preserving.
Intuitively speaking, we can under-approximate the greatest lower bound of clean
instances by not enforcing some of the interacting MDs.
On the other side, we can over-approximate the least upper bound
by assuming that the matching functions are similarity preserving. This would lead
us to keep applying
MDs on the assumption that unresolved similarities still persist. We present two
chase-like procedures
to compute $D_{\downarrow}$ and $D_{\uparrow}$ corresponding to these approximations.

\subsubsection*{Under-approximating the greatest lower bound.}
To provide an under-approximation for the greatest lower bound of all clean instances,
we provide a new chase-like procedure, which enforces only MDs that are
enforced in every clean instance. These MDs are applicable
to those initial similarities that exist in the original dirty instance, which are never broken by
enforcing other MDs during any chase procedure of producing a clean instance.

Let $\Sigma$ be a set of MDs, and $\varphi, \varphi' \in \Sigma$. We say that
$\varphi$ {\em precedes} $\varphi'$ if the set of attributes on the left-hand side of $\varphi'$
contains the attribute on the right-hand side of $\varphi$. We say that $\varphi$
{\em interacts with} $\varphi'$ if there are MDs $\varphi_1,\ldots,\varphi_k \in
\Sigma$, such that $\varphi$ precedes $\varphi_1$, $\varphi_k$ precedes $\varphi'$,
and $\varphi_i$ precedes $\varphi_{i+1}$ for $i \in [1,k-1]$, i.e., the interaction relationship
can be seen as the transitive closure of precedence relationship.

Let $D_0$ be a dirty database instance.
Let $\varphi: R_1[X_1] \approx R_2[X_2] \rightarrow R_1[A_1] \rightleftharpoons
R_2[A_2]$ be an MD in $\Sigma$. We say $\varphi$ is {\em freshly applicable} on $t_1,t_2$ in
$D_0$ if $t_1^{D_0}[X_1] \approx t_2^{D_0}[X_2]$, and $t_1^{D_0}[A_1] \neq t_2^{D_0}[A_2]$. We say
$\varphi$ is {\em safely applicable} on $t_1,t_2$ in $D_0$ if $\varphi$ is freshly
applicable on $t_1,t_2$ in $D_0$, and for every $\varphi' \in \Sigma$ that interacts with
$\varphi$, $\varphi'$ is not freshly applicable on $t_1,t_3$ or $t_2,t_3$ in $D_0$ for any tuple
$t_3$ (see Example~\ref{ex:approximate}).

\begin{definition}
\label{def:under} \em
For an instance $D_0$ and a set of MDs $\Sigma$, an instance $D_k$ is
$(D_0,\Sigma)$-{\em under clean} if there exists a finite sequence of instances
$D_1,\ldots, D_{k-1}$, such that
\begin{enumerate}
\item For every $i \in [1,k]$, $(D_{i-1},D_i)_{[t_1^i,t_2^i]} \models \varphi^i$,
for some $\varphi^i \in \Sigma$ and tuple identifiers $t_1^i,t_2^i$, such that
$\varphi^i$ is safely applicable on $t_1^i,t_2^i$ in $D_0$.
\item For every MD $\varphi: R_1[X_1] \approx R_2[X_2] \rightarrow R_1[A_1] \rightleftharpoons
R_2[A_2]$ in $\Sigma$ and tuples $t_1,t_2$, such that
$\varphi$ is safely applicable on $t_1,t_2$ in $D_0$,
we have $t_1^{D_k}[A_1] = t_2^{D_k}[A_2]$. \boxtheorem
\end{enumerate}
\end{definition}
Definition~\ref{def:under} characterizes a chase-based procedure that keeps enforcing
MDs that are safely applicable in the original dirty instance until all such MDs are enforced.
Notice that an under clean instance may not be stable. Moreover, safely applicable MDs
never interfere with each other, in the sense that enforcing one of them never breaks
the initial similarities in the dirty instance that are needed for enforcing other safely
applicable MDs.

\begin{proposition}
\label{prop:uniqueunder} \em
For every instance $D_0$ and every set of MDs $\Sigma$, there is a unique
$(D_0,\Sigma)$-under clean instance $D_{\downarrow}$.
\boxtheorem
\end{proposition}

\ignore{
\comlb{Here we should emphasize/explain that the safe applicability of each
$\varphi^i$ is against
$D_0$, the original instance. Also, we have to say that the last $t_1,t_2$ are the
tuples in $D_0$
with the same tuple ids as $t_1^i, t_2^i$. (I guess that is what you want, right?) I
miss a termination condition for
the sequence. Could there be a $D_{k+1}$? I guess the answer is yes, but for a
reviewer it may be difficult
to follow. Maybe some comments in the paper along these lines would help. Also say
that a final
instance $D_k$ for which there are no safely applicable MDs may not be stable. Etc. }

\comlb{I guess here there is still room for (essential) non-determinism. I mean,
there might be several branches, not
only
one that that you can stop wherever you want. There might be non-comparable (w.r.t.
$\sqsubseteq$) under clean
instances. Right? Something should be said.}
\comsk{I believe there is a unique under clean instance, because the order enforcing safely
applicable MDs does not matter. I will check this.}

\comlb{Will the coming example illustrate the non-stability mentioned above?}
}

\noindent
Clearly, an under clean instance $D_{\downarrow}$ can be computed in polynomial
time in the size of the dirty instance $D_0$. To construct it, we first need to identify safely
applicable MDs in $D_0$, and then enforce them in any arbitrary order until no such MDs
can be enforced. Next we show that $D_{\downarrow}$ is an under-approximation to
every $(D_0,\Sigma)$-clean instance. Intuitively, this is because $D_{\downarrow}$
is obtained by enforcing MDs that are enforced in every chase-based procedure of producing
a clean instance.

\begin{proposition}
\label{prop:sound} \em
({\em Soundness of under-approximation})
For every $(D_0,\Sigma)$-under clean instance $D_{\downarrow}$ and
every $(D_0,\Sigma)$-clean instance $D$, we have $D_{\downarrow} \sqsubseteq D$.
\boxtheorem
\end{proposition}

\noindent
Notice that an arbitrary $(D_0,\Sigma)$-clean instance $D$ may not be a sound under-approximation
for every other $(D_0,\Sigma)$-clean instances $D'$, because $D \sqsubseteq D'$ may not hold.

\ignore{
\comlb{Maybe the missing example could show the possible incompleteness here.}
}

Let $D_{\downarrow}$ be a $(D_0,\Sigma)$-under clean instance.
Then from Propositions~\ref{prop:sound} and \ref{prop:monotone},
we immediately obtain the following result.

\begin{theorem}
\em
For every monotone query $Q$, we have \linebreak
$\Q(D_0) \sqsubseteq \Q(D_{\downarrow}) \sqsubseteq {\it Cert}_{\Q}(D_0)$. \boxtheorem
\end{theorem}

\begin{example}
\label{ex:approximate} \em
Consider the instance $D_0$ and set of MDs $\Sigma$ in Example~\ref{nocycle-example}.
Observe that MD $\varphi_1$ is safely applicable on the first and second tuples in $D_0$.
Moreover, $\varphi_2$ is freshly applicable, but not safely applicable on the second and
third tuples. Accordingly, we obtain $(D_0,\Sigma)$-under clean instance $D_{\downarrow}$,
shown below, by enforcing $\varphi_1$ on the first two tuples.

\begin{center}
\begin{tabular}{c|ccc}
$D_{\downarrow}$ & $A$ & $B$ & $C$ \\ \hline
& $a_1$ & $\langle b_1, b_2 \rangle$ & $c_1$ \\
& $a_2$ & $\langle b_1, b_2 \rangle$ & $c_2$ \\
& $a_3$ & $b_3$ & $c_3$
\end{tabular}
\end{center}

\noindent
Notice that for the two $(D_0,\Sigma)$-clean instances $D_2,D'_3$ in Example~\ref{nocycle-example},
we have $D_{\downarrow} \sqsubseteq D_2$ and $D_{\downarrow} \sqsubseteq D'_3$.
Also notice that $D_{\downarrow}$ is not a stable instance. Now consider
the query $\Q: \pi_C(\sigma_{A=a_2} R)$. This query behaves monotonically for our purpose, because
the values of attribute $A$ are not changing by enforcing MDs. If we pose $\Q$ to $D_{\downarrow}$,
we obtain $\Q(D_{\downarrow}) = \{c_2\}$. Observe that
${\it Cert}_{\Q}(D_0) = \{\langle c_1,c_2 \rangle\}$, and thus
$\Q(D_{\downarrow})$ provides an under-approximation for ${\it Cert}_{\Q}(D_0)$.
This example also shows that an arbitrary clean instance, $D'_3$ here, may not
provide a sound approximation to certain answer since
$\Q(D'_3) = \{\langle c_1, c_2, c_3 \rangle\} \not \sqsubseteq {\it Cert}_{\Q}(D_0)$.
\boxtheorem
\end{example}

\subsubsection*{Over-approximating the least upper bound.}
To provide an over-approximation for the least upper bound of all clean instances,
we {\em modify} every similarity relation so that the corresponding matching function
becomes similarity preserving. For a similarity relation $\approx_A$ and the corresponding
matching function $\match_A$, we define $\approx_A^*$ as follows: For every $a,a' \in {\it Dom}_A$,
$a \approx_A^* a'$ iff there is $a'' \in {\it Dom}_A$, such that $a \approx_A a''$ and
$\match_A(a', a'') = a'$. Given a set of MDs $\Sigma$, we obtain $\Sigma^*$ by replacing
every similarity relation $\approx_A$ in the MDs by $\approx_A^*$.

\begin{definition}
\label{def:over} \em
For an instance $D_0$ and a set of MDs $\Sigma$, an instance $D_{\uparrow}$ is
$(D_0,\Sigma)$-{\em over clean} if $D_{\uparrow}$ is $(D_0,\Sigma^*)$-clean. \boxtheorem
\end{definition}

\noindent
By Proposition~\ref{prop:unique}, for every instance $D_0$ and set of MDs $\Sigma$, there is a
unique $(D_0,\Sigma)$-over clean instance $D_{\uparrow}$, and moreover, it can be
computed in polynomial time in the size of $D_0$. To construct $D_{\uparrow}$, we first
need to obtain $\Sigma^*$, as described above, and enforce MDs in $\Sigma^*$ in any arbitrary
order until we get a stable instance w.r.t. $\Sigma^*$.
Next we show that $D_{\uparrow}$ is an over-approximation for every $(D_0,\Sigma)$-clean instance.
Intuitively, this is because $D_{\uparrow}$ is obtained by enforcing (at least) all MDs that are present
in any chase-like procedure of producing a clean instance.

\begin{proposition}
\label{prop:complete} \em
({\em Completeness of over-approximation}) \linebreak For every $(D_0,\Sigma)$-over clean
instance $D_{\uparrow}$ and every $(D_0,\Sigma)$-clean instance $D$, we have
$D \sqsubseteq D_{\uparrow}$. \boxtheorem
\end{proposition}
Notice again that an arbitrary $(D_0,\Sigma)$-clean instance $D$ may not be an over-approximation
for every other $(D_0,\Sigma)$-clean instance $D'$, because $D' \sqsubseteq D$ may not hold.

Let $D_{\uparrow}$ be a $(D_0,\Sigma)$-over clean instance.
Then from Propositions~\ref{prop:complete} and \ref{prop:monotone},
we immediately obtain the following result.

\begin{theorem}
\em
For every monotone query $Q$, we have \linebreak
${\it Poss}_{\Q}(D_0) \sqsubseteq \Q(D_{\uparrow})$. \boxtheorem
\end{theorem}

\begin{example}
\label{ex:approximate2} \em
(Example~\ref{ex:approximate} continued.)
By assuming that old similarities hold after applying matching functions
(e.g., $\langle b_1,b_2 \rangle \approx^* b_3$), we obtain the $(D_0,\Sigma)$-over clean
instance $D_{\uparrow}$ shown below.

\begin{center}
\begin{tabular}{c|ccc}
$D_{\uparrow}$ & $A$ & $B$ & $C$ \\ \hline
& $a_1$ & $\langle b_1, b_2 \rangle$ & $\langle c_1, c_2, c_3 \rangle$ \\
& $a_2$ & $\langle b_1, b_2 \rangle$ & $\langle c_1, c_2, c_3 \rangle$ \\
& $a_3$ & $b_3$ & $\langle c_1, c_2, c_3 \rangle$
\end{tabular}
\end{center}

\noindent
Notice that for the two $(D_0,\Sigma)$-clean instances $D_2,D'_3$ in Example~\ref{nocycle-example},
we have $D_2 \sqsubseteq D_{\uparrow}$ and $D'_3 \sqsubseteq D_{\uparrow}$.
If we pose query $\Q: \pi_C(\sigma_{A=a_2} R)$ to $D_{\uparrow}$, we obtain
$\Q(D_{\uparrow}) = \{\langle c_1, c_2, c_3 \rangle\}$. Observe that
${\it Poss}_{\Q}(D_0) = \{\langle c_1, c_2, c_3 \rangle\}$, and thus
$\Q(D_{\uparrow})$ provides an over-approximation for ${\it Poss}_{\Q}(D_0)$.
It can be seen that an arbitrary $(D_0,\Sigma)$-clean instance, say $D_2$ for instance,
may not provide a complete approximation to possible answer since
${\it Poss}_{\Q}(D_0) \not \sqsubseteq \Q(D_2) = \{\langle c_1, c_2 \rangle\}$. \hspace{6mm}$\blacksquare$
\end{example}

\section{A Case for Swoosh's Entity\\Resolution}
\label{sec:swoosh}

\newcommand{\R}{\nit{Rec}}
\newcommand{\tr}{\nit{true}}
\newcommand{\appA}{\approx_{\{A\}}}
\newcommand{\matchA}{\match_{\{A\}}}
\newcommand{\dmda}{D^m\!\!\!\downarrow}

In \cite{BenjellounGMSWW09}, a generic conceptual framework for entity resolution is
introduced. It considers a general
match relation $M$, which is close to our similarity predicates $\approx$, and a
general merge function, $\mu$, which is
close to our $\match$ functions.
In this section we establish a connection between our MD framework and Swoosh.
However,  a full comparison is
problematic, for several
reasons, among them: (a) Swoosh works at the record (tuple) level, and we
concentrate on the attribute level. (b)
Swoosh does not use tuple identifiers and some tuples may be discarded at the end,
those that are dominated
by others in the instance. The main problem is (a). However, to ease the comparison,
we consider a particular
(but still general enough) case of Swoosh, namely the combination of the {\it union
case} with {\em merge domination}.
In the following we embed this case
of Swoosh into our MD framework, thus showing the power of the latter.

Although it is not explicitly said in \cite{BenjellounGMSWW09}, it is safe to say
that the conceptual framework is
applied to
ground tuples of a single relational predicate, say $R$, which are called {\em
records} there.
In consequence, $\nit{Rec}$ denotes the set of ground tuples of the form
$R(\bar{s})$. If the attributes of $R$
are $A_1, \ldots, A_n$, then the component
$s_i$ of $\bar{s}$ belongs to an underlying domain $\nit{Dom}_{A_i}$.

Relation $M$ maps $\nit{Rec} \times \nit{Rec}$ into $\{\nit{true},\nit{false}\}$. When
two tuples are similar and have to be merged, $M$ takes the value
$\nit{true}$.
Moreover, $\mu$ is a partial function from
$\nit{Rec} \times \nit{Rec}$ into $\nit{Rec}$. It produces the merge of the two
tuples into a single tuple, and
is defined only when $M$ takes the value $\nit{true}$.

Now, the union case for Swoosh arises when the merge function $\mu$ produces the
union of the records,
defined as the component-wise union of attribute values. This latter union makes
sense if the attribute values are
sets of values from an even deeper
data domain.

More precisely, for each of the $n$ attributes $A_i$ of $R$, we consider $n$
possibly denumerable domains $D_{\!A_i}$. (Repeated attributes in $R$
share the same domain, but it is conceptually simpler
to assume that attributes are all different.)
Each $D_{\!A_i}$ has a similarity relation $\approx_{A_i}$, which is
reflexive and symmetric.
Now, for each
 attribute $A_i$ of $R$, its domain becomes $\nit{Dom}_{\!A_i} := \cup_{k \in
\mathbb{N}}
\mc{P}^k(D{\!_{A_i}})$, where $k>0$ and $\mc{P}^k(D_{\!A_i})$ denotes the set of
subsets of $D_{\!A_i}$ of cardinality
$k$.
In consequence, the elements
of $\R$ are of the form $R(s_1, \ldots, s_n)$, with each $s_i$ being a set that
belongs to $\nit{Dom}_{\!A_i}$.
  An initial instance $D$, before any entity resolution, will be a finite subset of
$\R$, and each
  attribute value in a record, say $s_i$ for $A_i$,  will be
 a singleton of the form $\{a_i\}$, with $a_i \in D_{\!A_i}$.

The $\approx_{A_i}$ relation on $D_{\!A_i}$ induces a similarity relation
$\approx_{\{A_i\}}$ on
$\nit{Dom}_{\!A_i}$, as follows: $s_1 \approx_{\{A_i\}} s_2$ holds iff there exist
$a_1 \in s_1, a_2 \in s_2$ with $a_1
\approx_{A_i} a_2$. Each $\approx_{\{A_i\}}$
is reflexive and symmetric. ($s \approx_{\{A_i\}} s$, because there is $a \in s$ and
$\approx_{A_i}$ is reflexive; and
symmetry follows from the symmetry of $\approx_{A_i}$.)
We also consider matching functions $\match_{\{A_i\}}: \nit{Dom}_{\!A_i} \times
\nit{Dom}_{\!A_i} \rightarrow
\nit{Dom}_{\!A_i}$ defined
 by $\match_{\{A_i\}}(s_1, s_2) := s_1 \cup s_2$. The structures $\langle
\nit{Dom}_{\!A_i}, \approx_{\{A_i\}},
 \match_{\{A_i\}}\rangle$ have all the properties described in Sections
\ref{sec:back} and \ref{sec:matFcs}.

\begin{proposition} \label{prop:sw} \em
 Each matching function $\match_{\{A_i\}}$ is total, idempotent, commutative and
 associative. It is also similarity preserving w.r.t. the $\approx_{\{A_i\}}$
similarity relation.  \hfill $\blacksquare$
 \end{proposition}

\noindent
Now, based on \cite{BenjellounGMSWW09}, we are ready to define the ``union match and
merge case" for Swoosh. Consider
two elements of $\R$, say $r_1 = R(\bar{s}^1), r_2 = R(\bar{s}^2)$: (a) $M(r_1, r_2)
:= \tr$ iff for some $i$,  $s^1_i
\approx_{\{A_i\}} s^2_i$.
 (b) When  $M(r_1, r_2) := \tr$, $\mu(r_1, r_2) := R(\match_{\{A_1\}}(s^1_1,s^2_1),
\ldots,$
 $\match_{\{A_n\}}(s^1_n,s^2_n))$.

 Function $M$ is reflexive and commutative, which follows from the reflexivity and
symmetry of the $\appA$.
 From \cite[Prop. 2.4]{BenjellounGMSWW09} we obtain that the combination of $M$ and
$\mu$ has Swoosh's ICAR properties,
 namely:\footnote{We use the superscript $s$, for Swoosh, to distinguish them from
the properties listed in Section
 \ref{sec:matFcs}.}
\begin{itemize}
\item [$I^s\!\!:$] Idempotency:  $\forall r \in \nit{Rec}, M(r,r) \mbox{ holds, and
} \mu(r,r) = r$.
\item [$C^s\!\!:$] Commutativity:  $\forall r_1, r_2 \in \nit{Rec}, M(r_1,r_2)
\mbox{ iff } M(r_2,r_1)$. Also
$M(r_1,r_2) \mbox{ implies}$ $\mu(r_1,r_2) = \mu(r_2,r_1)$.
\item [$A^s\!\!:$] Associativity: \ $\forall r_1,r_2,r_3 \in \nit{Rec}$, if
$\mu(r_1,\mu(r_2,r_3))$ and
    $\mu(\mu(r_1,r_2),r_3)$ exist, then they are equal.
\item [$R^s\!\!:$] Representativity: $\forall r_1,r_2,r_3,r_4 \in \nit{Rec}$, if
$r_3 = \mu(r_1,r_2)$ and
    $M(r_1,r_4)$
holds, then $M(r_3,r_4)$ also holds.
\end{itemize}
Now, Swoosh framework with $M$ and $\mu$ on $\nit{Dom}_A$ can be reconstructed by
means of the following
set $\Sigma^S$ of MDs: For $1 \leq i,j \leq n$,
\begin{equation}
R[A_i] \ \approx_{\{A_i\}} R[A_i] \ \longrightarrow \ R[A_j] \rlh R[A_j].
\label{eq:mds}
\end{equation}
The RHS of (\ref{eq:mds}) has to be applied, as expected, with the  matching
functions $\match_{\{A_j\}}$.
From Propositions \ref{prop:unique} and \ref{prop:sw}, we obtain that there is a
single $(D,\Sigma^S)$-clean
instance $D^m$. Consistently with our MD framework, we will assume that records have
tuple identifiers. Actually, in order to make more clear the comparison between the two frameworks, in this section and for MDs, we will  use
explicit tuple ids. They will be positioned in the first, extra attribute of each relation.
When the MDs are applied, only the new version of a tuple is kept.

In the case of Swoosh, the application of $\mu$ generates a new, merged tuple, but the old ones may stay.
However, Swoosh applies a pruning process
based on an abstract domination partial order between records, $\preceq^S$. The framework
concentrates mostly on the {\em merge domination relation} $\leq$, which is
defined by:
\begin{equation}
r_1 \leq r_2 \ :\Longleftrightarrow \ M(r_1,r_2) = \tr \mbox{ and } \mu(r_1,r_2) =
r_2. \label{eq:mergeOrder}
\end{equation}
The $I^s\!C^s\!A^s\!R^s$ properties make $\leq$ a partial order
  with some pleasant and expected monotonicity properties \cite{BenjellounGMSWW09}.

According to Section \ref{sec:matFcs}, we may consider each of the partial orders
$\preceq_{\{A_i\}}$ on the
$\nit{Dom}_{\!A_i}$: \
$s \preceq_{\{A_i\}} s' \ :\Leftrightarrow \ \match_{\{A_i\}}(s,s') = s'$. They
induce a $\preceq$ relation on $\R$
(cf. Definition \ref{def:domin}).

\begin{proposition} \em \label{prop:coinc}
The general dominance relation $\preceq$ on $\R$ coincides with the
merge domination relation $\leq$ obtained from $M$ and $\mu$. \boxtheorem
\end{proposition}
Given a dirty instance $D$, it is a natural question to ask about the relationship between
the clean instance $D^m$ obtained under our approach, by enforcing the above MDs,
and the {\em entity resolution} instance $D^s$ obtained directly via Swoosh.
The entity resolution $D^s$ is defined in \cite[Def. 2.3]{BenjellounGMSWW09} through
the conditions: 1. $D^s \subseteq \bar{D}$. 2. $\bar{D} \leq D^s$. 3. $D^s$ is
$\subseteq$-minimal for the two previous
conditions. Here, the partial-order $\leq$ between instances is induced by the
partial order $\leq$ between records as
in Definition
\ref{def:domin}. Instance $\bar{D}$ is the {\em merge closure} of $D$, i.e., the
$\subseteq$-minimal instance that
includes $D$
and is closed under $M$: $r_1, r_2 \in \bar{D} \mbox{ and } M(r_1, r_2) = \tr \
\Rightarrow \mu(r_1,r_2) \in \bar{D}$.

Notice that, in order to obtain $D^m$, tuple identifiers are introduced
and kept, whereas under Swoosh, there are no tuple identifiers and new tuples are
generated (via $\mu$) and some
are deleted (those $\leq$-dominated by other tuples). In consequence, the elements
of $D$ and $D^m$ under the MD
framework are of the form
$R(t,s_1, \ldots, s_n)$, and those in $D$ and $D^s$ under Swoosh are  the records
$r$ of the form $R(s_1, \ldots, s_n)$.
Since
$t$ is a tuple identifier, for every $R(t,s_1, \ldots, s_n)$, $r(t)$ denotes the
record $R(s_1, \ldots, s_n)$.


\begin{proposition} \label{prop:coinc2} \em
(a) For every $r$ in $D^s$ there is a tuple in $D^m$ with tuple identifier $t$, such
that $r(t)= r$.\\
(b) For every tuple $t \in D^m$, there is a record $r \in D^s$, such that $r(t) \leq
r$. \boxtheorem
\end{proposition}

\section{Conclusions} \label{sec:concl}

The introduction of matching dependencies (MDs) in \cite{Fan08} has been a valuable
addition to
data quality and data cleaning research. They can be regarded as {\em data quality
constraints}
that are declarative in nature and are based on a precise model-theoretic semantics.
They are bound to play an important role in database research and practice, together
(and
in combination) with
classical integrity constraints.

In this work we have made several contributions to the semantics of matching
dependencies.
We have refined the original semantics introduced in \cite{FanJLM09}, addressing
some important
open issues, but we have also introduced into the semantic framework  the notion of
matching function. This is an important
ingredient in entity resolution since matching functions indicate how attribute
values have to be merged
or identified.

The matching functions, under certain natural assumptions,
induce lattice-theoretic structures in the attributes'
domains. We
also investigated their  interaction
with the similarity relations in those same domains. This led us to introduce a
partial order of domination
between instances. This allows us to compare them in terms of information content.
This same notion was then
applied to sets of query answers.

On the basis of all these notions, we defined the
class of clean instances for a given dirty instance. They are the
intended and admissible instances
that could be obtained after enforcing the matching dependencies. The clean
instances were defined by means
of a chase-like procedure that enforces the MDs, while not making unjustified
changes on other attribute
values. The chase procedure stepwise improves the information content as related to
the domination order.

The notion of clean answer to a query posed to the dirty database was defined as a
pair formed by a lower and an upper bound in terms of information content for the
query answers. In this context we studied the notion of
monotone query w.r.t. to the domination order and how to relax a query into a monotone
one that provides
more informative answer than the original one.

The domination-monotone relational query language introduced uses the
lattice-theoretic
structure of the domain, and is interesting in its own right. It
certainly deserves further
investigation,  independently of MDs. It is an interesting open question to explore
 its connections with querying databases over partially ordered domains, with
incomplete or partial information \cite{levene00,imielinski84,libkin98,levene99}.
with
query relaxation in general \cite{koudas06,minker92}, and with relational languages based
on similarity relations
\cite{mendelzon95}.

We addressed some  problems around the enforcement of a set of matching dependencies for purposes of
data cleaning based on the
original proposal of \cite{Fan08, FanJLM09}, by explicitly making use of matching
functions. We studied issues such as the existence and uniqueness of clean instances,
the computational
cost of computing them, and the complexity of computing clean answers.
We identified cases where clean query answering is tractable, e.g., when there is a
single clean instance. However, we established that this problem is intractable in general.
We proposed polynomial time approximations.

Identifying other tractable cases and more efficient approaches to the intractable
ones is  part of ongoing research.  We are currently investigating the use
of logic programs with stable model semantics in the specification of clean
instances and in clean query answering. This idea has been
investigated in consistent query answering and has led to useful insights and
implementations \cite{ArenasBC03,greco03,BarceloBB01,eiter08,caniup10}. This route
would avoid the explicit computation of the clean instances, and clean query answering
could be done on top of the
program. However, the lattice-theoretic structure of the domains and the domination
order create a scenario that is substantially
different from the one encountered in database repairs w.r.t. classical integrity constraints.




\appendix

\section{Some Proofs}

\defproof{Lemma~\ref{lemma:lattice}}{
(1) Let $D$ be the instance $D_1 \curlyvee D_2$.
Clearly, $D_1 \sqsubseteq D$ and $D_2 \sqsubseteq D$.
Now let $D'$ be an arbitrary instance such that $D_1 \sqsubseteq D'$ and
$D_2 \sqsubseteq D'$, and let $t$ be a tuple in $D$. Then, by definition, $t$ is in $D_1$
or in $D_2$, and hence there should be a tuple $t'$ in $D'$ such that $t^D \preceq t'^{D'}$.
Therefore, we have $D \sqsubseteq D'$, and thus $D$ is the least upper bound of $D_1,D_2$.

\noindent
(2) Let $t$ be the tuple $t_1 \curlywedge t_2$. Clearly, $t \preceq t_1^{D_1}$ and $t \preceq t_2^{D_2}$.
Let $t'$ be an arbitrary tuple such that $t' \preceq t_1^{D_1}$ and $t' \preceq t_2^{D_2}$.
Then $t'[A] \preceq t_1^{D_1}[A]$ and $t'[A] \preceq t_2^{D_2}[A]$ for every attribute $A$ in the schema.
Thus, $t'[A] \preceq \glb(t_1^{D_1}[A],t_2^{D_2}[A])$ for every attribute $A$, and hence
$t' \preceq t$.

\noindent
(3) Let $D$ be the instance $D_1 \curlywedge D_2$. Let $t$ be a tuple in $D$.
Then there exist tuples $t_1$ in $D_1$ and $t_2$ in $D_2$, such that $t = t_1 \curlywedge t_2$,
and thus $t^D \preceq t_1^{D_1}$ and $t^D \preceq t_2^{D_2}$. Therefore, it
holds $D \sqsubseteq D_1$ and $D \sqsubseteq D_2$.

Let $D'$ be an arbitrary instance such that $D' \sqsubseteq D_1$ and $D' \sqsubseteq D_2$,
and let $t'$ be a tuple in $D'$. Then there exist tuples $t_1$ in $D_1$ and $t_2$ in $D_2$,
such that $t'^{D'} \preceq t_1^{D_1}$ and $t'^{D'} \preceq t_2^{D_2}$, and thus
$t'^{D'} \preceq \glb(t_1^{D_1},t_2^{D_2})$, which exists in $D$. We thus
have $D' \sqsubseteq D$.
}

\defproof{Theorem \ref{finite-prop}}{(sketch)
It is easy to see that for every $i \in [1,k]$, we have $D_{i-1} \sqsubseteq D_{i}$ and
$D_{i} \not \sqsubseteq D_{i-1}$. That is, $D_i$ strictly dominates $D_{i-1}$ (for simplicity,
assume that the new generated tuples are not completely dominated by other tuples).
Consider a database instance $D_{\!{\it max}}$ that has a single tuple in every relation,
for which the value of every attribute $A$ is the result of matching all values of $A$
(and other attributes comparable with $A$) in the active domain of $D_0$. Clearly,
$D_{\!{\it max}}$ provides an upper bound for the instances in the sequence, and thus
the sequence stops after finite number of steps. Furthermore, the number of matching
applications needed to reach $D_{\!{\it max}}$ is polynomial in the size of $D_0$.}

\defproof{Proposition~\ref{prop:unique}}{
Let $D,D'$ be two $(D_0,\Sigma)$-clean instances. We use two lemmas.

\begin{lemma} \em
\label{lem:preserve}
Let $\match_A$ be a similarity preserving function, and $a_1,a_2,a_3,a_4$
be values in the domain ${\it Dom}_A$, such that $a_1 \preceq a_3$ and
$a_2 \preceq a_4$. If $a_1 \approx_A a_2$, then $a_3 \approx_A a_4$.
\end{lemma}

\begin{lemma}
\label{lem:unique} \em
Let $D_1, \ldots, D_k$ be a sequence of instances
such that $D=D_k$, and for every $i \in [1,k]$, $(D_{i-1},D_i)_{[t_1^i,t_2^i]} \models \varphi^i$,
for some $\varphi^i \in \Sigma$ and tuple identifiers $t_1^i,t_2^i$.
Then for every $i \in [0,k]$, the following holds:
\begin{enumerate}
\item $t_1^{D_i}[A_1] \preceq t_1^{D'}[A_1]$, for every tuple identifier $t_1$ and every attribute $A_1$.
\item  if $t_1^{D_i}[A_1] \approx t_2^{D_i}[A_2]$, then $t_1^{D'}[A_1] \approx t_2^{D'}[A_2]$,
for every two tuple identifiers $t_1,t_2$ and two comparable attributes $A_1,A_2$.
\end{enumerate}
\end{lemma}
We prove Lemma~\ref{lem:unique} by an induction on $i$. For $i=0$, we clearly have
$t_1^{D_0}[A_1] \preceq t_1^{D'}[A_1]$ since $D'$ is a $(D_0,\Sigma)$-clean instance.
Moreover, if $t_1^{D_0}[A_1] \approx t_2^{D_0}[A_2]$, then $t_1^{D'}[A_1] \approx t_2^{D'}[A_2]$
by Lemma~\ref{lem:preserve}.

Suppose $1$ and $2$ hold for every $i < j$. If $1$ holds for $i=j$, then $2$ also holds for $i=j$ by
Lemma~\ref{lem:preserve}. Suppose $1$ does not hold for $i=j$:
$t_1^{D_j}[A_1] \not \preceq t_1^{D'}[A_1]$. Since $1$ holds for every $i < j$, the value of
$t_1^{D_j}[A_1]$ should be different from $t_1^{D_{j-1}}[A_1]$. Therefore, there should be an MD
$\varphi^j: R_1[X_1] \approx R_2[X_2] \rightarrow R_1[A_1] \rightleftharpoons R_2[A_2]$ in $\Sigma$
and a tuple identifier $t_2$, such that $D_j$ is the immediate result of enforcing $\varphi^j$ on $t_1,t_2$
in $D_{j-1}$. That is, $t_1^{D_{j-1}}[X_1] \approx t_2^{D_{j-1}}[X_2]$,
$t_1^{D_{j-1}}[A_1] \neq t_2^{D_{j-1}}[A_2]$, and $t_1^{D_j}[A_1] = t_2^{D_j}[A_2] =
\match_A(t_1^{D_{j-1}}[A_1], t_2^{D_{j-1}}[A_2])$. Since $t_1^{D_{j-1}}[X_1] \approx t_2^{D_{j-1}}[X_2]$,
by induction assumption, we have $t_1^{D'}[X_1] \approx t_2^{D'}[X_2]$, and thus,
$t_1^{D'}[A_1] = t_2^{D'}[A_2]$, because $D'$ is a stable instance.
Again by induction assumption, $t_1^{D_{j-1}}[A_1] \preceq t_1^{D'}[A_1]$
and $t_2^{D_{j-1}}[A_2] \preceq t_2^{D'}[A_2] = t_1^{D'}[A_1]$. Therefore,
$t_1^{D_j}[A_1] = \match_A(t_1^{D_{j-1}}[A_1], t_2^{D_{j-1}}[A_2]) \preceq t_1^{D'}[A_1]$
since $\match_A$ takes the least upper bound, which leads to a contradiction.

To prove the first part of Proposition~\ref{prop:unique}, notice that, from Lemma~\ref{lem:unique},
we obtain $t_1^D[A_1] \preceq t_1^{D'}[A_1]$ and $t_1^{D'}[A_1] \preceq t_1^D[A_1]$
for every tuple identifier $t_1$ and every attribute $A_1$. Thus, the two $(D_0,\Sigma)$-clean
instances $D,D'$ should be identical.

To prove the second part of the proposition, let
$\varphi: R_1[X_1] \approx R_2[X_2] \rightarrow R_1[A_1] \rightleftharpoons R_2[A_2]$
be an MD in $\Sigma$. By Lemma~\ref{lem:unique}, if $t_1^{D_0}[X_1] \approx t_2^{D_0}[X_2]$,
then $t_1^D[X_1] \approx t_2^D[X_2]$, for every two tuple identifiers $t_1,t_2$.
Since $D$ is a stable instance, $t_1^D[A_1] = t_2^D[A_2]$, and thus $(D_0,D) \models \varphi$
and $(D_0,D) \models \Sigma$.
}

\defproof{Proposition~\ref{prop:nointeraction}}{
Let $D,D'$ be two $(D_0,\Sigma)$-clean instances. It is enough to prove a lemma similar to
Lemma~\ref{lem:unique}.

\begin{lemma}
\label{lem:nointeraction} \em
Let $D_1, \ldots, D_k$ be a sequence of instances
such that $D=D_k$, and for every $i \in [1,k]$, $(D_{i-1},D_i)_{[t_1^i,t_2^i]} \models \varphi^i$,
for some $\varphi^i \in \Sigma$ and tuple identifiers $t_1^i,t_2^i$.
Then for every $i \in [0,k]$, it holds
\begin{enumerate}
\item $t_1^{D_i}[X_1] = t_1^{D_0}[X_1]$ and $t_2^{D_i}[X_2] = t_2^{D_0}[X_2]$, where $X_1,X_2$
are lists of attributes on the left-hand side of $\varphi^i$.
\item $t_1^{D_i}[A_1] \preceq t_1^{D'}[A_1]$, for every tuple identifier $t_1$ and every attribute $A_1$.
\end{enumerate}
\end{lemma}
Notice that $1$ trivially holds: since MDs are interaction free, there is no MD $\varphi \in \Sigma$, such that
the attributes on the right-hand side of $\varphi$ has an intersection with  $X_1,X_2$, and therefore no MD
enforcement could change the values in $t_1^{D_i}[X_1]$ or $t_2^{D_i}[X_2]$ into something different from the
original values in $D_0$.

We prove $2$ by an induction on $i$. For $i=0$, we clearly have
$t_1^{D_0}[A_1] \preceq t_1^{D'}[A_1]$ since $D'$ is a $(D_0,\Sigma)$-clean instance.
Now suppose $2$ holds for $i < j$, and it does not hold for $i=j$:
$t_1^{D_j}[A_1] \not \preceq t_1^{D'}[A_1]$. Then there should be an MD
$\varphi^j: R_1[X_1] \approx R_2[X_2] \rightarrow R_1[A_1] \rightleftharpoons R_2[A_2]$ in $\Sigma$
and a tuple identifier $t_2$, such that $D_j$ is the immediate result of enforcing $\varphi^j$ on $t_1,t_2$
in $D_{j-1}$. That is, $t_1^{D_{j-1}}[X_1] \approx t_2^{D_{j-1}}[X_2]$,
$t_1^{D_{j-1}}[A_1] \neq t_2^{D_{j-1}}[A_2]$, and $t_1^{D_j}[A_1] = t_2^{D_j}[A_2] =
\match_A(t_1^{D_{j-1}}[A_1], t_2^{D_{j-1}}[A_2])$.
Since $t_1^{D_{j-1}}[X_1] \approx t_2^{D_{j-1}}[X_2]$,
by part $1$ we have $t_1^{D'}[X_1] \approx t_2^{D'}[X_2]$, and thus
$t_1^{D'}[A_1] = t_2^{D'}[A_2]$, because $D'$ is a stable instance.
By induction assumption, $t_1^{D_{j-1}}[A_1] \preceq t_1^{D'}[A_1]$
and $t_2^{D_{j-1}}[A_2] \preceq t_2^{D'}[A_2] = t_1^{D'}[A_1]$. Therefore,
$t_1^{D_j}[A_1] = \match_A(t_1^{D_{j-1}}[A_1], t_2^{D_{j-1}}[A_2]) \preceq t_1^{D'}[A_1]$,
since $\match_A$ takes the least upper bound, which leads to a contradiction.
}

\defproof{Theorem \ref{theo:coNP}}{Consider relation schema $R(C, V, L)$, query $\Q: \pi_L(R)$, and set $\Sigma$ consisting
of two MDs $\varphi_1: R[C] \approx R[C] \rightarrow R[C] \rightleftharpoons R[C]$
and $\varphi_2: R[CV] \approx R[CV] \rightarrow R[L] \rightleftharpoons R[L]$.
The domains of attributes, similarity relations, and matching functions are as follows:
${\it Dom}_C = \{\bot, c, c_1,d_1, c_2, d_2 \ldots\}$, ${\it Dom}_V = \{\bot, y, x_1, x_2, \ldots\}$,
${\it Dom}_L = \{\bot, \top, +, - \}$. For every $c_i,d_i \in {\it Dom}_C$,
$c_i \approx d_i$ and $\match_C(c_i,d_i) = \match_C(d_i, c_i) = c$.
We also have $\match_L(+,-) = \match_L(-,+) = \top$. Notice that similarity
relations and match functions are not fully described here. The full descriptions
can be derived using the reflexivity and symmetry of similarity relations and
idempotency, commutativity, and associativity of match functions.

To prove membership in coNP, it is easy to see that given a certificate, which
is a $(D_0,\Sigma)$-clean instance $D$, we can check whether $\top \not \in \Q(D)$
in polynomial time.
To prove hardness, we reduce from 3SAT.
Let $C_1 \land \ldots \land C_N$ be CNF formula, where each clause $C_i$, $i \in [1,N]$,
is a disjunction of three literals $l_{i1} \lor l_{i2} \lor l_{i3}$, and each literal $l_{ik}$,
$k \in [1,3]$, is either $x_j$ or $\lnot x_j$ for some variable $x_j \in {\it Dom}_V$.
We create an instance $D_0$ of $R$ as follows. For every clause $C_i$ and
every literal $l_{ik}$ of variable $x_j$ in $C_i$, there is a tuple $t$ with
$t^{D_0}[C] = c_i$, $t^{D_0}[V] = x_j$, $t^{D_0}[L] = +$ if $l_{ik} = x_j$ (a positive literal),
and $t^{D_0}[L] = -$ if $l_{ik} = \lnot x_j$ (a negative literal). Moreover, for every clause $C_i$,
there is another tuple $t$ with $t^{D_0}[C] = d_i$, $t^{D_0}[V] = y$, and $t^{D_0}[L] = +$.

We show that the CNF formula $C$ is satisfiable if and only if
$\top \not \in {\it Cert}_{\Q}(D_0)$. Let $C$ be a
satisfiable formula. For each clause $C_i$, we pick a tuple corresponding to the
literal that is made true in the satisfying assignment and also the only tuple with
$t^{D_0}[C] = d_i$, and enforce the MD $\varphi_1$ on these two tuples. It is easy
to see that the result would be a stable instance $D$. In particular, $(D,D) \models \varphi_2$
since for each variable the satisfying assignment has picked only one of the positive or
negative literals to be true. Therefore, we do not need to enforce $\varphi_2$, which
means that $\top$ does not appear for any value of attribute $L$, and hence
$\top \not \in \Q(D)$, and $\top \not \in {\it Cert}_{\Q}(D_0)$.

Conversely, if $\top \not \in {\it Cert}_{\Q}(D_0)$, there is a $(D_0,\Sigma)$-clean
instance $D$ in which $\top$ does not appear for any value of attribute $L$. To obtain
the clean instance $D$ starting from $D_0$, we need to enforce $\varphi_1$ once for each clause $C_i$,
as described above, before we can enforce $\varphi_2$ on any tuple corresponding to $C_i$.
Moreover, for every two tuples in $D$ that match the left-hand side of $\varphi_2$, we should have
identical values for attribute $L$ (either $+$ or $-$), otherwise we would get $\top$
when enforcing $\varphi_2$. Therefore, for each clause, we can make true the literal corresponding
to the tuple on which $\varphi_1$ has been enforced, and obtain a correct satisfying assignment.
}

\defproof{Proposition~\ref{prop:monotone}}{
For every instance $D' \in \mathcal{D}$, we clearly have
$\glb_\sqsubseteq \{ D \mid D \in \mathcal{D} \} \sqsubseteq D'$, since $\Q$ is a monotone query,
it holds $\Q(\glb_\sqsubseteq \{ D \mid D \in \mathcal{D} \}) \sqsubseteq \Q(D')$.
Consequently,
$Q(\glb_\sqsubseteq \{ D \mid D \in \mathcal{D} \}) \sqsubseteq \glb_\sqsubseteq \{ \Q(D') \mid D' \in \mathcal{D} \}$.
With a similar argument, we can show that the second equation of Proposition~\ref{prop:monotone} holds.
}

\defproof{Proposition~\ref{prop:relaxedra}}{(sketch)
We can prove the proposition by an structural induction on the relational
algebra expression.
It is enough to show that every operation in $\mc{RA}_{\preceq}$ is monotone.
Projection, Cartesian product, and union are clearly monotone operators w.r.t. $\sqsubseteq$.
Now let $D, D'$ be two instances such that $D \sqsubseteq D'$. Consider query
$\Q: \sigma_{a \preceq A} R$ for relation $R$ in the schema.
Let $t$ be an $R$-tuple in $\Q(D)$. Clearly $t$ is an $R$-tuple
in $D$. Therefore, there is an $R$-tuple $t'$ in $D'$ with $t \preceq t'$. Now it holds
$a \preceq t^D[A] \preceq t'^{D'}[A]$, and thus $t'$ is in $\Q(D')$.

Now consider the query $\Q': \sigma_{A_1 \Join_{\preceq} A_2} R$, and let
$t$ be an $R$-tuple in $\Q'(D)$. Then there is $a \in {\it Dom}_A$ s.t. $a \preceq t^D[A_1]$,
$a \preceq t^D[A_2]$, and $a \neq \bot$. Since $D \sqsubseteq D'$,
there should be an $R$-tuple $t'$ in $D'$ with $t \preceq t'$.
Now it holds $a \preceq t^D[A_1] \preceq t'^{D'}[A_1]$
and $a \preceq t^D[A_2] \preceq t'^{D'}[A_2]$.
Therefore, $t'$ is in $\Q(D')$.

}

\defproof{Proposition~\ref{prop:uniqueunder}}{
The proof of this proposition is very similar to that of Proposition~\ref{prop:nointeraction}.
Let $D_{\downarrow},D'_{\downarrow}$ be two $(D_0,\Sigma)$-under clean instances.
It is enough to prove the following lemma.

\begin{lemma}
\label{lem:uniqueunder} \em
Let $D_1, \ldots, D_k$ be a sequence of instances for deriving $D_{\downarrow}$
as described in Definition~\ref{def:under}. Then for every $i \in [0,k]$, it holds
\begin{enumerate}
\item $t_1^{D_i}[X_1] = t_1^{D_0}[X_1]$ and $t_2^{D_i}[X_2] = t_2^{D_0}[X_2]$, where $X_1,X_2$
are lists of attributes on the left-hand side of $\varphi^i$.
\item $t_1^{D_i}[A_1] \preceq t_1^{D'_{\downarrow}}[A_1]$, for every tuple identifier $t_1$
and every attribute $A_1$.
\end{enumerate}
\end{lemma}
Suppose that for some $i \in [0,k]$, $t_1^{D_i}[X_1] \neq t_1^{D_0}[X_1]$. Then there exists $j < i$,
tuple $t_3$, and MD $\varphi^j \in \Sigma$, such that $(D_{j-1},D_j)_{[t_1,t_3]} \models \varphi^j$,
with attribute $B_1 \in X_1$ on the right-hand side of $\varphi^j$. MD $\varphi^j$ has to be safely
applicable on $t_1,t_3$ in $D_0$, which means that $\varphi^i$ cannot be safely applicable on $t_1,t_2$
in $D_0$, a contradition. The proof of part 2 is similar to the proof of part 2 in Lemma~\ref{lem:nointeraction}.
}

\defproof{Proposition~\ref{prop:sound}}{
Let $D_{\downarrow}$ be a $(D_0,\Sigma)$-under clean instance, and
$D$ be a $(D_0,\Sigma)$-clean instance.
The proof follows from the following two lemmas.

\begin{lemma} \em
For every two tuples $t_1,t_2$ and MD
$\varphi: R_1[X_1] \approx R_2[X_2] \rightarrow R_1[A_1] \rightleftharpoons R_2[A_2]$
in $\Sigma$, such that $\varphi$ is safely applicable on $t_1,t_2$ in $D_0$, it holds
$t_1^D[X_1] = t_1^{D_0}[X_1]$ and $t_2^D[X_2] = t_2^{D_0}[X_2]$.
\end{lemma}

\begin{lemma} \em
Let $D_1, \ldots, D_k$ be a sequence of instances for deriving $D_{\downarrow}$
as described in Definition~\ref{def:under}. Then for every $i \in [0,k]$, it holds
$t_1^{D_i}[A_1] \preceq t_1^D[A_1]$, for every tuple identifier $t_1$
and every attribute $A_1$.
\end{lemma}
The proof of lemma is by induction on $i$ and is very similar to the proof
part 2 in Lemma~\ref{lem:nointeraction}.

}

\defproof{Proposition \ref{prop:sw}}{In fact: If $s_1 \approx_{\{A\}} s_2$, then there are $a_1 \in s_1, a_2 \in s_2$ with $a_1 \approx_A a_2$. Since $a_2$ also
  belongs to $s_2 \cup s_3$, for every $s_3 \in \nit{Dom}_A$, it holds $s_2 \cup s_3 = \match_{\{A\}}(s_2,s_3) \approx_{\{A\}} s_1$.}

  \defproof{Proposition \ref{prop:coinc}}{For $\preceq_{\{A\}}$ on $\nit{Dom}_A$ it holds:
$s \preceq_{\{A\}} s' \ :\Leftrightarrow \ \matchA(s,s') = s' \ \Leftrightarrow \ s \cup s' = s' \ \Leftrightarrow \
s \subseteq s'$. Now, for records $r_1 = R(s_1^1,\ldots,s_n^1), r_2 = R(s_1^2, \ldots, s_n^2),$ it holds $r_1 \preceq r_2  :\Leftrightarrow  \mbox{ for every } i, s_i^1 \preceq_{\{A\}} s_i^2 \Leftrightarrow  \mbox{ for every } i,
s_i^1 \subseteq s_i^2$.

On the other side, from (\ref{eq:mergeOrder}) we obtain that, for records $r_1 = R(s_1^1,\ldots,s_n^1), r_2 = R(s_1^2, \ldots, s_n^2)$, it holds:
$r_1 \leq r_2  \Leftrightarrow  M(r_1,r_2) = \tr \mbox{ and for every } i, s_i^1 \subseteq s_i^2$. Since the $s_i^j$ are non-empty, the first condition
on the RHS is implied by the second one.}

\defproof{Proposition \ref{prop:coinc2}}{(sketch) As a preliminary and useful remark, let us mention that the
$I^s\!C^s\!A^s\!R^s$ properties make $\leq$
a partial order
  with the following monotonicity properties
   \cite{BenjellounGMSWW09}: (A) $M(r_1, r_2) = \tr \ \Longrightarrow \ \ r_1 \leq
\mu(r_1,r_2) \mbox{ and}$ $r_2 \leq
   \mu(r_1,r_2)$. (B) $ r_1 \leq r_2 \mbox{ and } M(r_1,r) = \tr \ \Longrightarrow \
M(r_2,r) = \tr$.
(C) $ r_1 \leq r_2 \mbox{ and } M(r_1,r) = \tr \ \Longrightarrow \ \mu(r_1,r) \leq
\mu(r_2,r)$.
(D) $r_1 \leq s, \ r_2 \leq s \mbox{ and } M(r_1,r_2) = \tr \ \Longrightarrow \
\mu(r_1,r_2) \leq s$.

More specifically for our proof, first notice that every application of $\mu$ can be
simulated by a finite sequence of
enforcement of the MDs in (\ref{eq:mds}).
More precisely, given two tuples $R(t_1,\bar{s}^1), R(t_2,\bar{s}^2)$ in an instance
$D$, such that
$M(r(t_1),$ $r(t_2))$ holds, then  $\mu(r(t_1),r(t_2)) = r(t)$ for some  tuple
$R(t,$ $r(t))$ of the form
$\match_{\{A_{i_1}\}} \cdots \match_{\{A_{i_n}\}}(R(t_1,\bar{s}^1),
R(t_2,\bar{s}^2))$, i.e., obtained
by enforcing the MDs. Furthermore, it holds $r(t_1)$ $\leq r(t)$ and $r(t_2) \leq r(t)$.

Conversely, every enforcement of an MD in (\ref{eq:mds}) is dominated by a tuple
obtained
through the application of $\mu$. More precisely, for tuples $R(t_1,\bar{s}^1),
R(t_2,\bar{s}^2)$ in an instance $D$
for which $s_j^1 \approx_{\{A_j\}} s_j^2$ holds, it also holds $M(t_1(r), t_2(r))$,
and
$\match_{\{A_j\}}(R(t_1,\bar{s}^1), R(t_2,\bar{s}^2)) \leq R(t,\mu(t_1(r), t_2(r))$
for some tuple id $t$ (actually,
$t_1$ or $t_2$).

Now, for (a), consider $D^m\!\!\!\downarrow \ :=\{r(t)~|~ R(t,\bar{s}) \in D^m\}$
(from where duplicates are
eliminated).
It is good enough to prove that $D^s \subseteq D^m\!\!\!\downarrow$. For this it
suffices to prove that $\dmda$
satisfies conditions 1. and 2. on the entity resolution instance, namely:
1. $\dmda \ \subseteq \bar{D}$ and  2. $\bar{D} \leq \dmda$. The first condition
follows from the definition
(or construction) of $D^m$ as a stable instance obtained by minimally applying the
MDs and when justified only.
The second condition follows from the simulation and properties of $\mu$ as a
finitely long enforcement
of the MDs.

Now (b) follows from the domination of a tuple obtained by applying one MD by a
tuple obtained applying $\mu$ as
described above.
}

\end{document}